\documentclass[
 reprint,
 amsmath,amssymb,
 aps,
]{revtex4-2}
\usepackage{amsmath,amsfonts,amssymb,amsbsy}
\usepackage{mathrsfs,latexsym}
\usepackage{graphicx}
\usepackage{xcolor}
\usepackage{subcaption}
\usepackage{dcolumn}
\usepackage[bookmarks,
bookmarksopen = true,
bookmarksnumbered = true,
linktocpage,
colorlinks = true,
linkcolor = blue,
urlcolor  = blue,
citecolor = blue,
anchorcolor = green,
hyperindex = true,
hyperfigures]{hyperref}
\usepackage{array}
\usepackage{booktabs}
\usepackage{multirow}
\usepackage{placeins}
\usepackage{blindtext}
\usepackage{appendix}
\usepackage{fontawesome} 
\usepackage{bm}
\usepackage{times}
\usepackage{float}
\usepackage{wrapfig}
\usepackage{physics}
\usepackage[utf8]{inputenc} 
\usepackage[T1]{fontenc} 
\usepackage{amsmath,amssymb}

\newcommand{\ben}{\begin{enumerate}}
\newcommand{\een}{\end{enumerate}}
\newcommand{\bit}{\begin{itemize}}
\newcommand{\eit}{\end{itemize}}
\newcommand{\beq}{\begin{equation}}
\newcommand{\eeq}{\end{equation}}
\newcommand{\bsa}{\begin{subequations}\begin{eqnarray}}
\newcommand{\esa}{\end{eqnarray}\end{subequations}}
\newcommand{\bea}{\begin{eqnarray}}
\newcommand{\eea}{\end{eqnarray}}
\newcommand{\bean}{\begin{eqnarray*}}
\newcommand{\ean}{\end{eqnarray*}}

\usepackage{xcolor}

\preprint{WUB/25-04\\CERN-TH-2025-114}

\begin{document}

\begin{flushleft}
\textbf{WUB/25-04, CERN-TH-2025-114}
\end{flushleft}
\title{S-wave flavor-singlet meson mixing in QCD with light and charm quarks}%

\author{Juan Andrés Urrea-Niño}
\affiliation{School of Mathematics, Trinity College Dublin, Dublin 2, Ireland}
\affiliation{Department of Physics, Bergische Universit\"at Wuppertal, Gau{\ss}str. 20, 42119 Wuppertal, Germany}
\author{Roman~H\"ollwieser} 
\author{Francesco~Knechtli} 
\author{Tomasz~Korzec} 
\affiliation{Department of Physics, Bergische Universit\"at Wuppertal, Gau{\ss}str. 20, 42119 Wuppertal, Germany}

\author{Jacob Finkenrath}
\affiliation{CERN, Esplanade des Particules 1, 1211 Geneva 23, Switzerland}

\author{Michael Peardon}
\affiliation{School of Mathematics, Trinity College Dublin, Dublin 2, Ireland}


\begin{abstract}
We investigate the mixing between flavor-singlet light meson and charmonium operators in the S-wave channels, i.e. pseudo-scalar and vector channel, at two different pion masses. We measure statistically significant non-zero correlations between operators with different quark content corresponding to off-diagonal entries of a flavor-singlet mixing correlation matrix. By solving a GEVP we extract the low-lying energy spectrum and compare it with the one obtained by the different types of operators separately. We also calculate the overlaps between the states created by different operators and the energy eigenstates of the theory and find that all types of operators contribute to resolve the states of interest.
\end{abstract}

\maketitle


\section{\label{sec:intro}Introduction}

Quantum Chromodynamics (QCD) is the sector of the standard model describing the strong force which binds quarks and gluons inside hadrons. The strong interactions confine these constituents, which are never observed directly in experiment. There are still theoretical and experimental questions about confinement to be answered and a major focus of interest are the unexpected $XYZ$ states discovered by the Belle and BaBar experiments at the start of the new millennium (see Ref.~\cite{Brambilla:2019esw, Godfrey:2008nc} for a review). To make matters even more interesting, a recent study proposed the $X(2370)$ as a candidate for a pseudo-scalar glueball \cite{AblikimGlueball}. Glueballs remain a fundamental open question of QCD which still awaits definitive experimental and theoretical confirmation. Lattice QCD provides a framework to study such exotic states, as well as conventional ones, from first principles. A system of particular interest as a starting point for studying exotic states is charmonium. A rich spectrum has been observed including candidate hybrids in recent
calculations \cite{Liu:2012ze,Cheung:2016bym}, however these use an approximation where states are assumed to be stable.  In precision studies of
low-lying states \cite{DeTar:2018uko,Hatton}, one significant remaining uncertainty is
the omission of charm quark-line disconnected (charm annihilation) diagrams.  These effects are
suppressed by the Okubo-Zweig-Iizuka (OZI) rule \cite{Close:1979bt} but this
dynamics is poorly understood outside of a perturbative regime.  An example of this is the shift of the mass of the pseudo-scalar particle, the $\eta_c$, having the opposite sign in NRQCD perturbation theory compared to recent indirect lattice QCD calculations \cite{Hatton}. Many of the $XYZ$ resonances appear very close
to the open-charm threshold and determining their nature will require the
delicate dynamics occurring at thresholds to be investigated precisely. Furthermore, charmonium has an energy scale close to the pseudo-scalar glueball and their mixing induced by charm annihilation tests the validity of the OZI rule beyond perturbation theory.\\

We have previously developed an improvement of the technique to compute smeared all-to-all quark propagators called distillation \cite{Peardon:2009gh}. By introducing a profile in distillation space \cite{Knechtli2022, Urrea-Nino:2022gne, Nino:2021klm} the effective masses reach their plateau values (the masses of the particles) at significantly earlier Euclidean time separations between source and sink of the meson. This allowed us to compute the contribution of charm annihilation to the mesonic correlation functions \cite{Knechtli2022} and to get a first signal for the mixing of charmonium with glueballs \cite{Urrea-Nino:2022gne}. In this work we aim to extend these calculations to systematically study the mixing between flavor-singlet hadron creation operators in $N_f = 3 + 1$ QCD. Examples for such operators are charmonium $\Bar{c}c$ and flavor-singlet light mesons. While there are studies which explore flavor mixing in an $N_f = 2 + 1$ setup with degenerate up/down quarks and a strange quark \cite{Dudek:2011tt}, the inclusion of the charm quark is fundamental to reach the energy scale relevant for the pseudo-scalar glueball. Non-zero mixing correlations between these meson and gluonic operators have been previously observed \cite{Urrea-Nino:2022gne, Zhang2022, Jiang2023}. This indicates that the energy eigenstates which overlap with states created by meson operators also overlap with states created by gluonic operators, hinting at possible mixing between meson and glueball states. In the light sector and in the presence of $SU(3)$ light flavor symmetry, the pion and $\eta^{\prime}$ masses are distinguished due to the effects of disconnected contributions to the temporal two-point function of the $\eta^{\prime}$, so these cannot be neglected. For the $\eta_c$ this disconnected contribution, while in principle necessary to build the correlation function, has been shown to lead to a relatively small shift in the mass \cite{Levkova, McNeile,Forcrand,Hatton}. This has led to charm disconnected contributions being often neglected in lattice calculations. However, on the lattice these two particles, together with all their radial excitations, lie within the same symmetry channel and therefore the correlation functions built from charmonium or light meson operators are expected to be dominated by a common ground state at large enough time separations. This is also true for any flavor-singlet operator with the correct quantum numbers, such as purely gluonic ones. In this work we focus on measuring this mixing between the different types of meson operators with pseudo-scalar and vector quantum numbers by including all relevant disconnected contributions to correlation functions. By doing so, we can quantify how well each different type of operator can sample the energy eigenstates from the light meson energy scale up to the charmonium one. This can shed some light on the dynamics which occurs solely due to quark-anti-quark annihilation and provides a more complete approach towards the study of the finite volume energy spectrum in the flavor-singlet channel. Such a calculation is a fundamental first step towards more scattering studies of resonances, one of which could be the elusive glueball state.

\section{\label{sec:methods}Methods}

\subsection{\label{sec:dist}Improved Distillation}
Improved distillation was introduced in ~\cite{Nino:2021klm,Urrea-Nino:2022gne, Knechtli2022} as a modification to the distillation technique~\cite{Peardon:2009gh} which uses distillation profiles as an additional degree of freedom in the construction of hadronic creation operators. These profiles were introduced at quark level and optimized at meson level for different symmetry channels via the Generalized Eigenvalue Problem (GEVP) formulation~\cite{Luscher:1990ck, Blossier:2009kd}. The starting point is a meson operator of the form $\bar{q}\Gamma q$, where $\Gamma$ fixes the quantum numbers. After Wick contractions, the two-point temporal correlation function at zero spatial momentum can be written in terms of \textit{perambulators} 
\begin{align}
    \tau[t_1,t_2] &= V[t_1]^{\dagger} D^{-1} V[t_2],
\end{align}
where the columns of $V[t]$ are labeled as $V_{i,\alpha}[t]$ and correspond to the $i$-th 3D Laplacian eigenvector of time slice $t$, denoted as $v_i[t]$, placed in time $t$ and Dirac index $\alpha$, i.e
\begin{align}
    \left[ V_{i,\alpha}[t]\right]_{a,\vec{x},\beta,t^{\prime}} &= \delta_{\alpha \beta} \delta_{t t^{\prime}} \left[v_i[t]\right]_{a,\vec{x}},
\end{align}
where $a$ is the color index, as well as elementals
\begin{equation}
\Phi[t]= V[t]^{\dagger} \Gamma V[t].
\end{equation}
The introduction of different quark distillation profiles amounts to using multiple elementals where the contributions from different eigenvectors are weighted differently, i.e
\begin{equation}
\Phi^{(k)}[t]_{\substack{i j\\ \alpha \beta}} = g_{k}(\lambda_{i}[t])^{*}v_{i}[t]^{\dagger} \Gamma_{\alpha \beta} v_{j}[t] g_k(\lambda_j[t]),
\end{equation}
where $g_k(\lambda)$ is a quark distillation profile and $\lambda_i[t]$ is the eigenvalue associated with $v_i[t]$. We consider $N_b$ meson operators $\mathcal{O}_k$ based on different quark distillation profiles and build a correlation matrix with entries
\begin{equation}
C_{kl}(t) = \left \langle \mathcal{O}_{k}(t) \mathcal{O}_{l}^\dagger(0)  \right \rangle,
\end{equation}
where $\left \langle ... \right \rangle$ denotes the expectation value integral. By solving the GEVP
\begin{equation}
C(t) \mathbf{w}_n(t,t_G) = \rho_n(t,t_G) C(t_G) \mathbf{w}_n(t,t_G)
\end{equation}
we can access the energy of the $n$-th energy eigenstate from the large-time asymptotics of the generalized eigenvalues
\begin{equation}
\lim_{t\rightarrow\infty} \rho_n(t,t_G) = C_n e^{-E_n (t-t_G)},
\end{equation}
where $C_n$ is a time-independent constant. We can also use the generalized eigenvectors $\mathbf{w}_n(t,t_G)$ to define a linear combination of the original $N_b$ meson operators which creates a state with the largest overlap with the $n$-th energy eigenstate:
\begin{equation}
\Tilde{\mathcal{O}}_n(t) = \sum_{k=1}^{N_b} \mathbf{w}_n^{(k)}(t_R,t_G) \mathcal{O}_k(t),
\end{equation}
where $t_R$ is a suitably chosen reference time. The elemental corresponding to this operator is given by
\begin{equation}
\tilde{\Phi}^{(n)}[t]_{\substack{i j\\ \alpha \beta}} = v_{i}[t]^{\dagger} \Gamma_{\alpha \beta} v_{j}[t] \; \Tilde{f}_n(\lambda_i[t],\lambda_j[t]),
\end{equation}
where the optimal meson distillation profile $\Tilde{f}_n(\lambda_i[t],\lambda_j[t])$ is a linear combination of products of the form $g_k\left( \lambda_i[t] \right) g_k\left( \lambda_j[t]  \right)$ which depends on the GEVP vectors. The numerical stability of the GEVP depends strongly on how linearly independent the basis operators $\mathcal{O}_k$ are and the use of different profiles does not necessarily guarantee the absence of near-degeneracies. To counter this issue, we use the pruning procedure presented in~\cite{Balog, Niedermayer}. There, we calculate the $N_p < N_b$ singular vectors $u_a$, $a=1,...,N_p$, corresponding to the largest singular values of $C(t)$ at reference time $t_p$ and define a projected correlation matrix
\begin{equation}
\tilde{C}_{ab}(t) = u_a^{\dagger} C(t) u_b
\end{equation}
which we then use for the GEVP. Since the correlation matrix is hermitian and positive definite up to statistical noise, the left and right singular vectors are equal. This pruning not only keeps the contributions from orthogonal operators but also removes the ones whose contributions are small and more sensitive to the statistical noise. It also means the operator basis consists now of linear combinations of the original $N_b$ operators. Since our basis of operators involves ones with different flavor content we must be careful when pruning to avoid mixing operators we wish to keep separate. We elaborate on this later on.

\subsection{Mixing correlation matrix}
In this work we consider $N_f = 3 + 1$ flavors and therefore charmonium and light mesons must be treated differently due to flavor symmetry. For charmonium, a meson operator with fixed quantum numbers $\bar{c} \Gamma c$ leads to a correlation function given by
\begin{align}
C_{cc}(t) &= - \left \langle \text{Tr}\left( \Phi[t] \tau_c[t,0] \overline{\Phi}[0] \tau_c[0,t]   \right) \right \rangle_{\text{gauge}}\\
&+ \left \langle  \text{Tr}\left( \Phi[t] \tau_c[t,t] \right) \text{Tr}\left( \overline{\Phi}[0] \tau_c[0,0] \right)  \right \rangle_{\text{gauge}} \nonumber,
\end{align}
where $\overline{\Phi}[t]$ is the elemental corresponding to $\overline{\Gamma} = \gamma_0 \Gamma^{\dagger} \gamma_0$ and $\tau_c[0,t]$ is the charm perambulator. The first term will be referred to as the connected contribution while the second term is the disconnected one. The latter has higher variance in Monte Carlo estimates and is considerably more difficult to calculate than the former, however it is necessary if we want to study mixing between flavor singlet states. For light mesons, the correlation function depends on the quark content of the meson in terms of the up, down and strange degenerate quarks of this theory. The neutral pion, a member of the pseudo-scalar octet, with quantum numbers $0^{-+}$ has a flavor content $\frac{1}{\sqrt{2}}(\bar{u}u - \bar{d}d)$. The resulting correlation function for a suitable choice of $\Gamma$ is
\begin{equation}
C_{\pi}(t) = - \left \langle  \text{Tr}\left( \Phi[t] \tau_l[t,0] \overline{\Phi}[0] \tau_l[0,t] \right)  \right \rangle_{\text{gauge}},
\end{equation}
which is equal to the one of the octet $\eta$ with flavor content $\frac{1}{\sqrt{6}}(\bar{u}u + \bar{d}d - 2\bar{s}s)$ and to the other six octet correlators when the three flavors are degenerate in mass. A flavor singlet has content $\frac{1}{\sqrt{3}}(\bar{u}u + \bar{d}d + \bar{s}s)$, which results in the correlation function
\begin{align}
C_{ll}(t) &= - \left \langle  \text{Tr}\left( \Phi[t] \tau_l[t,0] \overline{\Phi}[0] \tau_l[0,t] \right)  \right \rangle_{\text{gauge}}\\
&+ 3\left \langle  \text{Tr}\left( \Phi[t] \tau_l[t,t] \right) \text{Tr}\left( \overline{\Phi}[0] \tau_l[0,0] \right)  \right \rangle_{\text{gauge}} \nonumber,
\end{align}
which also has a connected and disconnected contribution. Charmonium and flavor singlet light mesons with the same quantum numbers are in the same symmetry channel and therefore the temporal correlation between them can be non-zero and both operators create states with non-zero overlaps onto the energy eigenstates of the fixed symmetry channel. More importantly, the ground state that dominates the spectral decompositions of light meson and charmonium correlations is the same for both types. The mixing correlation between the charm and light sector is given by
\begin{equation}
C_{lc}(t) = \sqrt{3}\left \langle  \text{Tr}\left(  \Phi[t] \tau_c[t,t] \right) \text{Tr}\left( \overline{\Phi}[0] \tau_l[0,0]  \right)  \right \rangle_{\text{gauge}},
\end{equation}
which further emphasizes disconnected contributions are essential for mixing studies. This mixing can be included in a mixing correlation matrix of the form
\begin{equation}
C(t) = 
\begin{pmatrix}
C_{cc}(t) & C_{cl}(t) \\
C_{lc}(t) & C_{ll}(t).
\end{pmatrix}
\end{equation}
The presence of non-zero off-diagonal entries $C_{cl}(t)$ and $C_{lc}(t)$ is known as "explicit mixing" \cite{Bali:2005fu}. Nonetheless, flavor-mixing can also happen  implicitly in the $C_{cc}(t)$ and $C_{ll}(t)$ matrices. This happens via the disconnected contributions, as light quark loops can contribute to correlations which only involve charm quarks in the operators and vice versa. Since this mixing is not explicitly introduced via an off-diagonal entry in $C$, the name "implicit mixing" is suitable. For our meson operators we use $\Gamma = \gamma_0 \gamma_5$ for the pseudo-scalar and $\Gamma = \gamma_i$ for the vector channel.  We do not use $\Gamma = \gamma_5$ the flavor-singlets since it can give rise to finite volume effects related to topological quantities \cite{BaliArtifacts, UrbachArtifacts, AokiArtifacts, Jiang2023, OttnadArtifacts}. When working with a diverse basis of operators to build a correlation matrix we need to be careful with numerical issues that might arise which are related to how linearly independent the operators are from each other as well as with their different normalizations. While the previously mentioned pruning would clearly help with this issue, it would also change our operators into a basis where light meson and charmonium ones are mixed. To improve the conditioning of our correlation matrix while keeping light meson and charmonium operators separated we perform what we call \textit{partial} pruning. Here, we calculate the singular vectors corresponding to the largest singular values of the charmonium and light mesons diagonal blocks at time $t_p$, denoted as $V_{c}$ and $V_{l}$, taking into account only the connected correlations. If the effects of the disconnected pieces are small then this new basis should be a good starting point. If these effects are large, then this new basis is simply a different choice than the original and no significant physical information should be lost. We define a partially pruned correlation matrix as
\begin{align}
\hat{C}(t) &=
\begin{pmatrix}
V_c^{\dagger} & 0\\
0 & V_l^{\dagger} 
\end{pmatrix}
\begin{pmatrix}
C_{cc}(t)  & C_{cl}(t)\\
C_{lc}(t) & C_{ll}(t) 
\end{pmatrix}
\begin{pmatrix}
V_c & 0 \\
0 & V_l 
\end{pmatrix}
\label{eq:PartialPruning}
\end{align}
Once the partially pruned correlation matrix is built, we normalize it using
\begin{align}
    \tilde{C}_{ij}(t) = \frac{\hat{C}_{ij}(t)}{\sqrt{\hat{C}_{ii}(a)\hat{C}_{jj}(a)}},
\end{align}
so that all diagonal correlators are normalized to unity at $t=a$. We use this partially pruned, normalized matrix in our GEVP analysis. Unless stated otherwise, we use $t_p = t_G = a$. We are aware that we should ideally choose a larger value of $t_G$ as presented in \cite{Blossier:2009kd} but the statistical accuracy of the correlators does not allow for it.\\

Finally, the statistical analysis is done using the \textit{pyerrors} library \cite{Joswig2023} which uses the $\Gamma$-method \cite{Wolff2004, Wolff2007,Schaefer2011} with automatic differentiation \cite{Ramos2019}. \\

\subsection{Mixing overlaps}
\label{subsec:MixingOverlaps}
While mixing of flavor-singlet creation operators is not expected a priori to be zero by any symmetry, it is often assumed that it is small in magnitude as it can only happen through disconnected loops for the case of one-particle operators considered in this work. However, even if small, this mixing can give us useful information about the composition of the energy eigenstates of the theory. Previous studies involving charmonium and glueball operators have approached this issue by fitting mixing correlations to carefully chosen forms based on assumptions on the type of states the creation operators create \cite{Zhang2022, Jiang2023}. In this work we measure mixing correlations between light meson and charmonium operators but quantify the mixing via the GEVP vectors and resulting spectrum. If $\bra{\Omega} \hat{\mathcal{O}}_{i} \ket{n} \bra{n} \hat{\mathcal{O}}_j^{\dagger} \ket{\Omega}$, the contribution from energy eigenstate $\ket{n}$ to the mixing off-diagonals 
is non-zero with statistical significance then the states $\hat{\mathcal{O}}_{\bar{c}c}^{\dagger} \ket{\Omega}$ and $\hat{\mathcal{O}}_{\bar{l}l}^{\dagger} \ket{\Omega}$ have non-zero contributions from different energy eigenstates. These overlaps however can still be small and therefore it is not guaranteed that the GEVP built from these operators can resolve all energy eigenstates, e.g multi-particle states could be missed if we only use single-particle creation operators. To access the value of the overlaps $\bra{n} \hat{\mathcal{O}}_i^{\dagger} \ket{\Omega}$ we use the approach described in \cite{Dudek2008} based on the GEVP eigenvectors. This approach has a phase ambiguity and therefore we will work with absolute values. Furthermore, each creation operator comes with an unknown renormalization factor, making it impossible to compare overlaps coming from different operators. To compare overlaps coming from a same operator we define 
\begin{align}
    \xi_i^{(n)} = \frac{|\bra{n} \hat{\mathcal{O}}_i^{\dagger} \ket{\Omega}|}{\sum_{m} |\bra{m} \hat{\mathcal{O}}_i^{\dagger} \ket{\Omega}|},
\end{align}
where the sum goes over the energy eigenstates of interest.

\section{\label{sec:ensembles}Lattice Ensembles}
We use two different $N_f = 3 + 1$ ensembles with non-perturbatively improved Wilson fermions and Lüscher-Weisz gauge action which are described in the following subsections. The main difference between them is the pion mass; one of them has a heavier pion mass than the other. Varying the pion mass is relevant to control the possible decay channels of the scalar glueball. By considering only the quantum numbers, this glueball can decay into even numbers of pions and the maximum number will depend on the pion mass. Based on the calculation of the quenched scalar glueball being around 1800 MeV \cite{Morningstar:1999rf}, the pion masses are chosen such that in the ensemble with the lighter pion this glueball can decay in up to four pions while in the ensemble with the heavier pion it can only decay into two. Parameters of both ensembles are summarized in Table \ref{table:Ensembles}.

\begin{center}
\begin{table}
\begin{tabular}{||c c c c c c c c||} 
 \hline
  & $\kappa_l$ & $\kappa_c$ & $a$ [fm] & $m_{\pi}$ [MeV] & $N_{\text{cnfg}}$ & $N_v^{l}$ & $N_v^{c}$\\ [0.5ex] 
 \hline\hline
 A1 & $0.134407$ & $0.12784$ & $0.05198(36)$ & $\approx 420$ & 9000 & 100 & 200\\ 
 \hline
 A1h & $0.13392$ & $0.12834$ & $0.0690(11)$ & $\approx 800$ & 2000 & 200 & 200\\
 \hline
\end{tabular}
 \caption{Ensemble details for this work. In both cases the lattice sizes are $96 \times 32^3$ and $\beta = 3.24$. $N_v^{l/c}$ is the number of distillation vectors used for light and charm perambulators respectively.}
 \label{table:Ensembles}
\end{table}
\end{center}

\subsection{\label{sec:A1l}Ensemble with lighter $m_{\pi}$}
For the measurements we use ensemble $A1$ of~\cite{Hollwieser:2020qri,Hollwieser:2019kuc}, which is a $96\times 32^3$ lattice with $m_{\pi} \approx 420$ Mev. The three light quarks are mass degenerate, with a mass close to the average light quark mass in nature. In addition the simulation contained a charm quark with a realistic mass. The $O(a)$ improvement of the Wilson action was specifically designed to reduce lattice artifacts in the presence of heavy quarks~\cite{Fritzsch:2018kjg}. Potential problems with the sampling of topological sectors, that typically plague simulations at fine lattice spacings,  were avoided entirely by using open boundary conditions in the time direction. It has been demonstrated, that the discretization effects are mild and that the volume is large enough for meson spectroscopy. Additionally, in \cite{Hollwieser:2019kuc} it was shown that this setup can reproduce physical charmonium well. The effect of slightly unrealistic light quark masses largely cancels between up/down quarks, that are too heavy and the strange quark, that is too light, in quantities that do not depend on the light quarks explicitly. The hadron masses in~\cite{Hollwieser:2020qri} were obtained using a traditional method with unsmeared operators, which restricted us to just the ground states in a few selected flavor-octet channels. 

\begin{figure}
    \centering
    \includegraphics[width=0.95\linewidth]{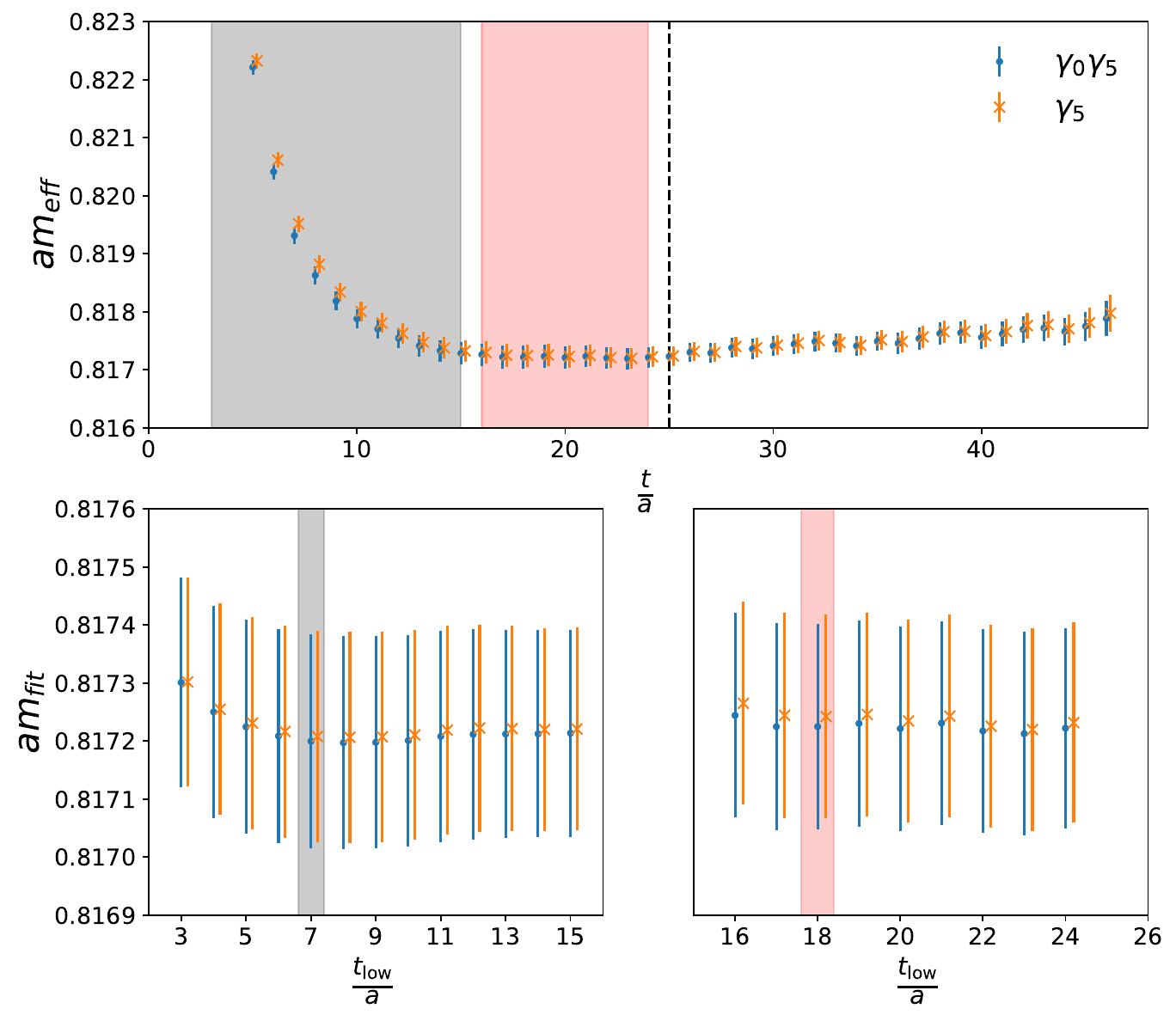}
    \caption{Plateau mass determination for the $\eta_c$ using only quark-connected correlation functions in ensemble A1. The lower left panel shows the result of the constant term in the constant + exponential fit as a function of $t_{low}$. The right panel corresponds to the same but for the constant fit. The points highlighted in the black and red shaded regions of the lower panels correspond to the values we take as the constant terms for these fits.}
    \label{fig:Fits_etaC_A1}
\end{figure}

\begin{figure}
    \centering
    \includegraphics[width=0.95\linewidth]{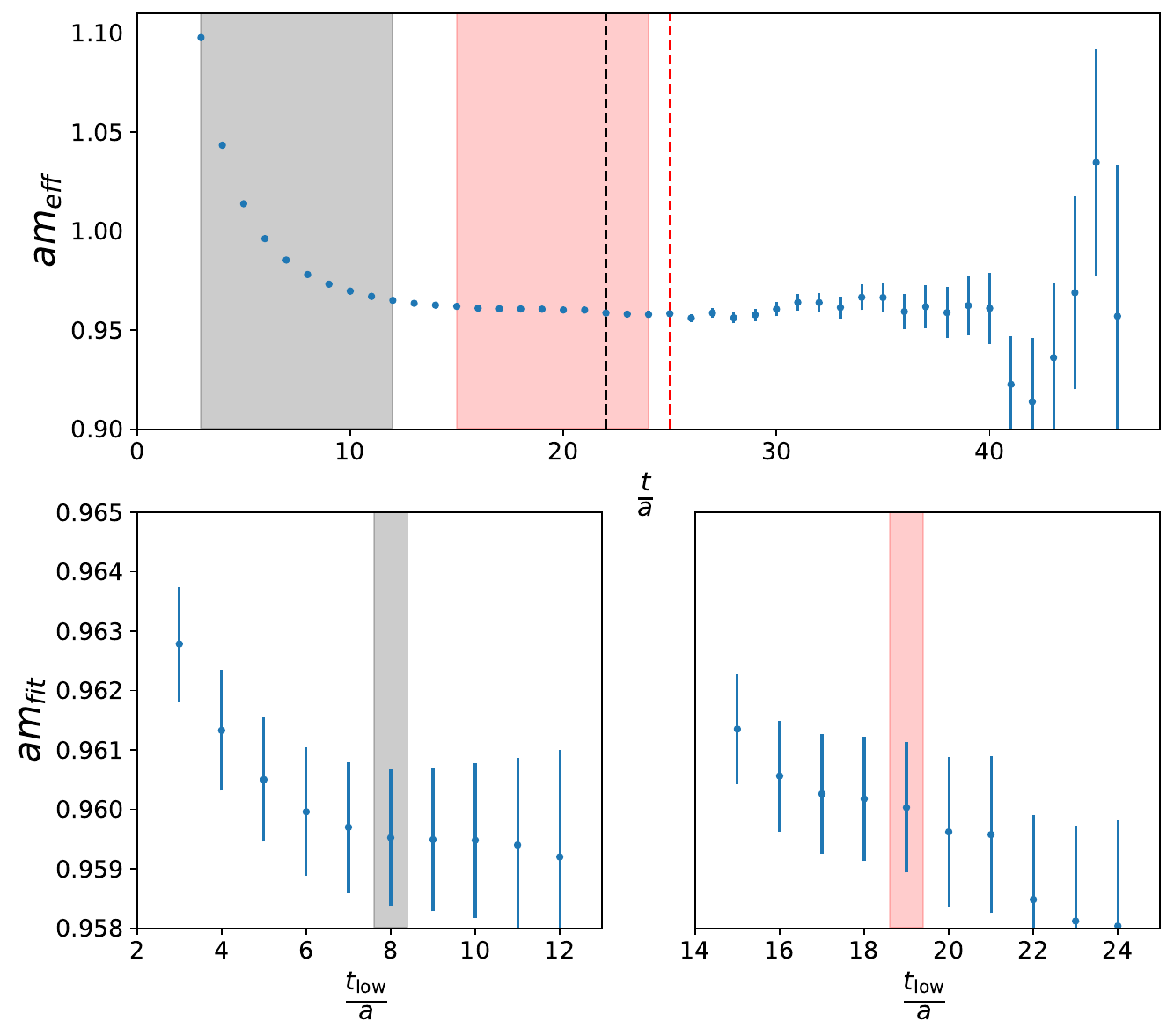}
    \caption{Plateau mass determination for the $h_c$ using only quark-connected correlation functions in ensemble A1. The panels are organized as in Fig. \ref{fig:Fits_etaC_A1}.}
    \label{fig:Fits_hc_A1}
\end{figure}

In total we use 9000 gauge configurations separated into two replica of 2000 configurations each and one replica with 5000. The latter replica has been generated with an efficient and stable Hessian-free force-gradient integrator \cite{Schafers:2024teg, Schafers:2025mgt} implemented in a modified version of the openQCD package \cite{Luscher:2012av}. We set the scale using the difference between the charmonium S- and P-wave spin singlets, i.e $\eta_c$ and $h_c$. For this we use only quark-connected correlators as they have the cleanest signal. To robustly extract the ground state plateau masses of these two channels we perform two different fits to the effective masses. In one we fit the data to a constant and in the other we fit it to a constant plus an exponential correction. We refer to Fig.~\ref{fig:Fits_etaC_A1} to aid in the explanation of these fits. For both fits we need to define a time interval where to perform them. In both cases we choose $t=25a$ to be the upper limit, shown in the upper panel by the vertical black dotted line. Since we work with open boundary conditions in time for the gauge field and only use perambulators in the bulk of the temporal extent, larger time separations have fewer time sources to average over and therefore statistical fluctuations are larger. Since our optimized operators allow us to access the ground state at early values of time, we take those for better fit stability. We try different values of the lower time limit of the fits, denoted as $t_{low}$, and the black shaded region contains all different values we use for the constant plus exponential fit while the red shaded region contains the ones used for the constant fit. We use the results for $\gamma_5$ and $\gamma_0 \gamma_5$ since we look at quark-connected correlations only, and their degeneracy shows there is no major difference in choosing one over the other in terms of suppressing excited state contamination. We choose one value per fit for both operators, shown in the lower panels of Fig. \ref{fig:Fits_etaC_A1}, fit these four points to a constant and take this to be the plateau value for the quark-connected $\eta_c$. We do the same for the $h_c$, except there we only use one operator ($\epsilon_{ijk}\gamma_j \gamma_k$) and the different fits have a different temporal upper limit. Fig.~\ref{fig:Fits_hc_A1} shows the fit regions we use, where the dashed lines mark the different upper limits. By taking the difference we measure in lattice units and taking the PDG value $541.27$ MeV \cite{ParticleDataGroup:2024cfk}, we obtain $a = 0.05198(36)$ fm, which is consistent with the value $a=0.0536(11)$ fm reported in \cite{Hollwieser:2020qri} calculated with a different method.

\subsection{\label{sec:A012h}Ensemble with heavier $m_{\pi}$}
We generated new lattice ensembles with three degenerate light quarks with a pion mass of $800$ MeV and a physical charm quark to close some decay channels for the scalar glueball. After tuning, the bare quark mass parameters are $\kappa_l = 0.13392$ and $\kappa_c = 0.12834$. To set the scale we calculate the mass difference between the charmonium S- and P-wave spin singlets, i.e $\eta_c$ and $h_c$, and match it to its physical value. In total we have 4000 configurations divided into two replica of 2000 configurations each, however we only sampled every second configuration, yielding a total of 2000 configurations for this study.

%
%
\section{Spectrum with flavor-singlet mixing}

\subsection{Pseudoscalar channel}

\begin{figure}[H]
\includegraphics[width=.95\linewidth]{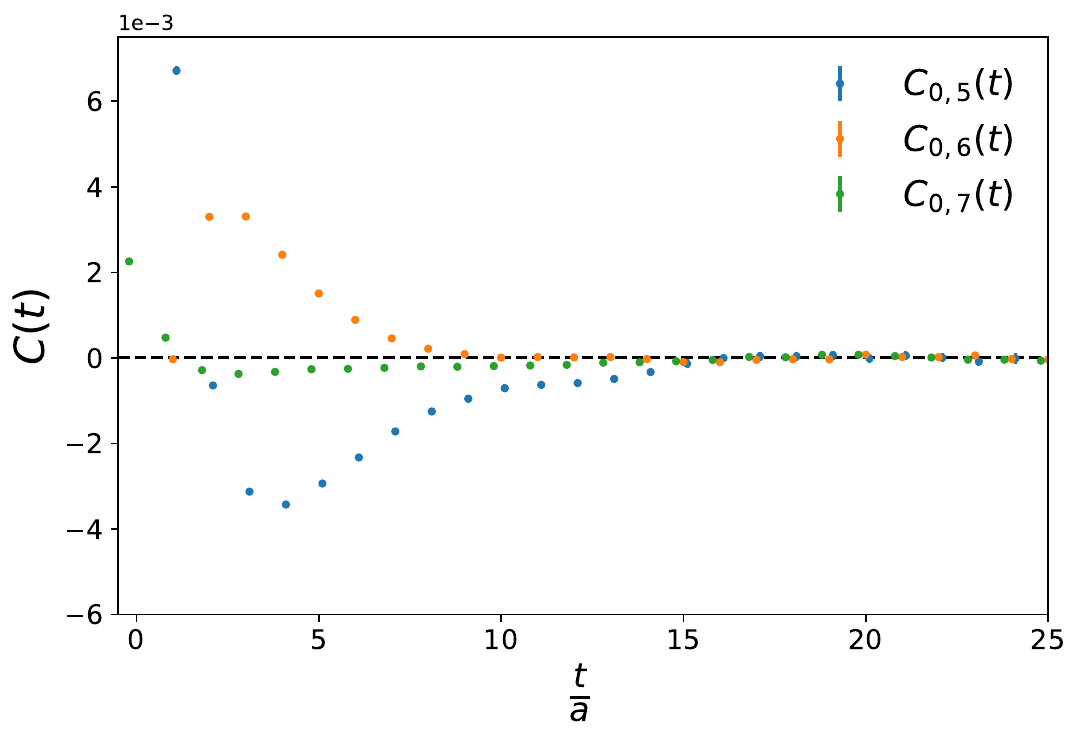}
\caption{Selected off-diagonal pseudoscalar correlations between the light meson and charmonium operators in ensemble A1. Indices $0,...,4$ label charmonium operators and $5,...,9$ light meson operators.}\label{fig:OffDiagonalCorrelations_G4G5_A1}
\end{figure}

\begin{figure}[H]
\includegraphics[width=.95\linewidth]{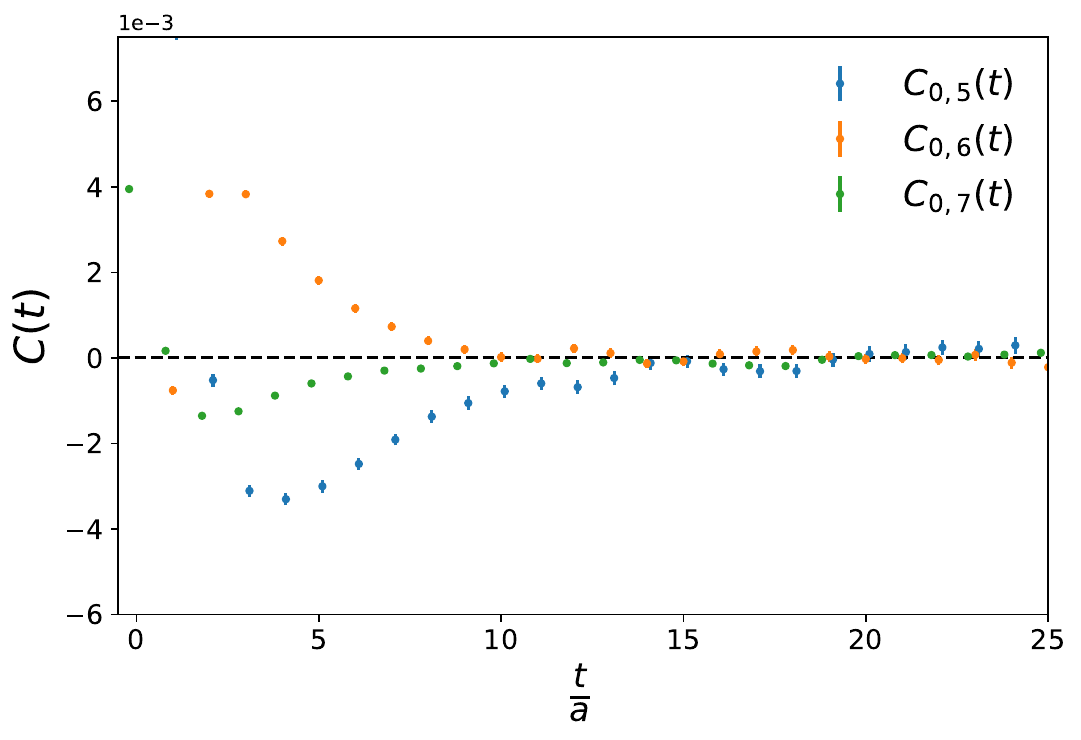}
\caption{Selected off-diagonal pseudoscalar correlations between the light meson and charmonium operators in ensemble A1-heavy. Indices $0,...,4$ label charmonium operators and $5,...,9$ light meson operators.}\label{fig:OffDiagonalCorrelations_G4G5_A1h}
\end{figure}

The starting point of our mixing analysis are the off-diagonal correlations between charmonium and light meson operators, some of which are plotted in Figs. \ref{fig:OffDiagonalCorrelations_G4G5_A1} and \ref{fig:OffDiagonalCorrelations_G4G5_A1h}. All correlations are non-zero with statistical significance at early times and some change sign. Since the spectral decomposition of off-diagonal correlations does not have explicit positivity, such a sign change is not forbidden. As the time separation increases, the exponential suppression of the different terms happens at different rates and therefore the result of subtractions can change sign. Additionally, some of these correlations grow in magnitude at small time separations before the exponential decay behavior eventually dominates. This is due to the same reason as the sign change: relative minus signs between different terms of the spectral decomposition.



\begin{figure}[H]
\includegraphics[width=.95\linewidth]{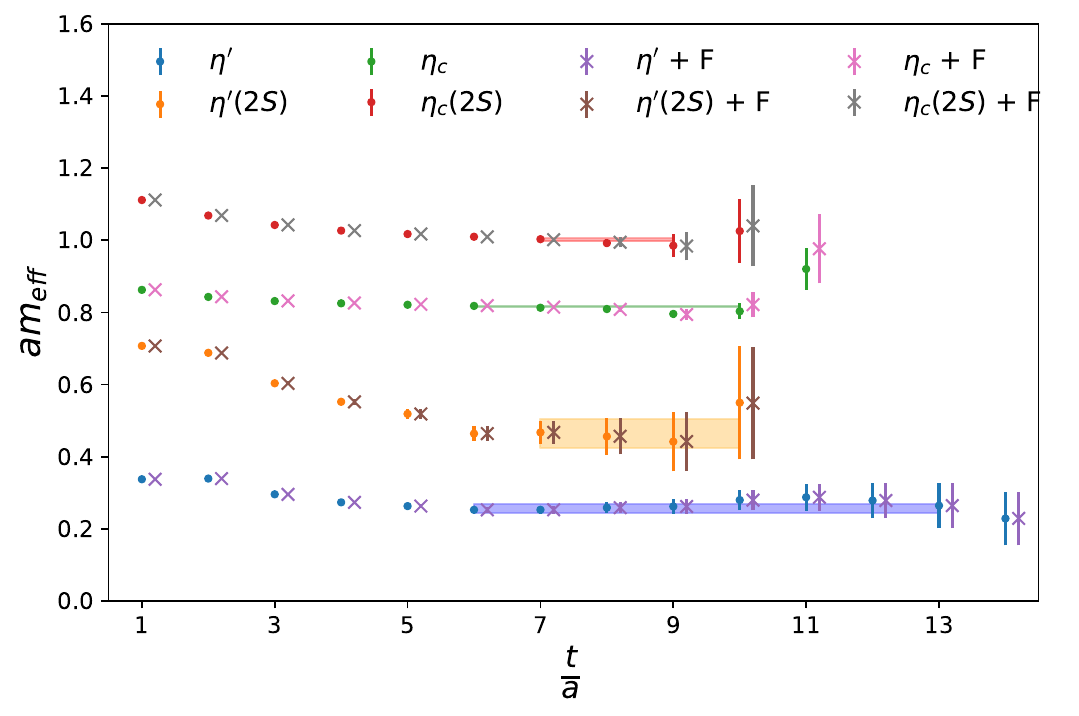}
\caption{Effective masses for the two lightest pseudoscalar light meson and charmonium states calculated with and without taking into account the mixing between both types of operators in ensemble A1.}\label{fig:A1mp_LightCharmG4G5_A1}
\end{figure}

\begin{figure}[H]
\includegraphics[width=.95\linewidth]{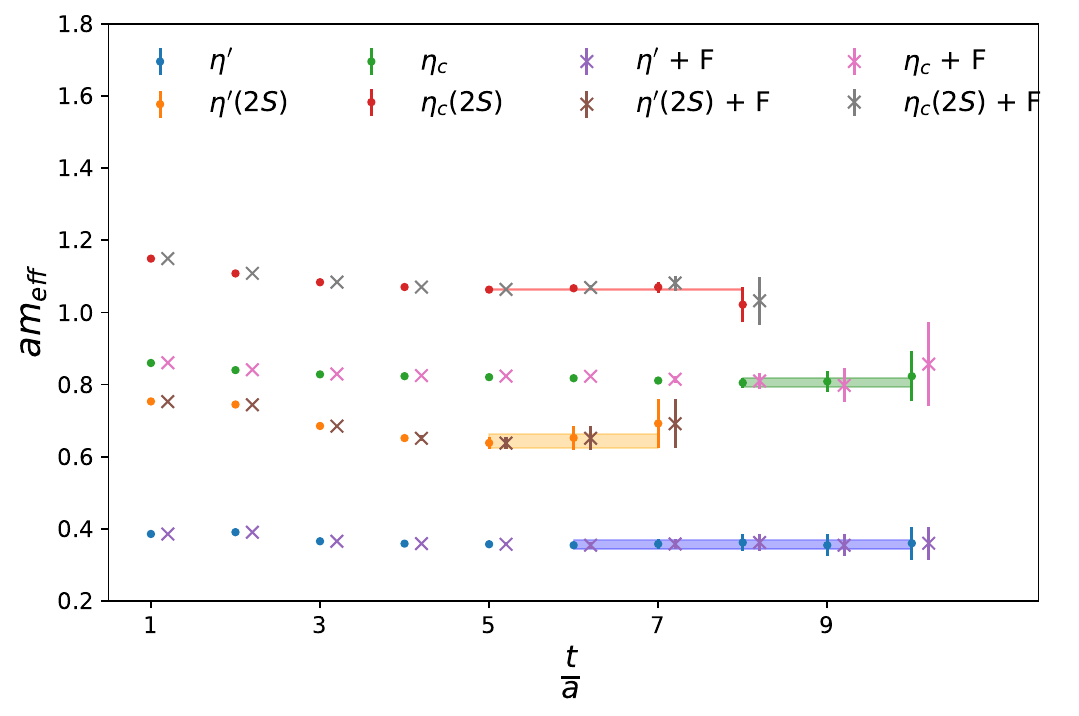}
\caption{Effective masses for the two lightest pseudoscalar light meson and charmonium states calculated with and without taking into account the mixing between both types of operators in ensemble A1-heavy.}\label{fig:A1mp_LightCharmG4G5_A1h}
\end{figure}

We first solve two separate GEVPs to establish a reference point to quantify the effects of the mixing: one with only light meson operators and one with only charmonium operators. From each of these we extract the ground and first excited states. We call them $\eta^{\prime}$ ($\eta_c$) and $\eta^{\prime}(2S)$ ($\eta_c (2S)$) respectively. This follows the experiment labeling, however it is important to keep in mind in the case of the charmonium-only GEVP, the physical $\eta_c$ is not the ground state, this would be the $\eta^{\prime}$. However, if the overlap between the states our operators create and light meson states is small enough then we can see an intermediate plateau around the mass of the true lightest charmonium state. Once we have these four reference states, we solve the GEVP using the matrix which accounts for the mixing between these two types of operators. Among the different states seen by this new GEVP, we take the four which are closest to our four reference states. Figs. \ref{fig:A1mp_LightCharmG4G5_A1} and \ref{fig:A1mp_LightCharmG4G5_A1h} show the effective masses of these different states. A ``+F'' appended to one of the four labels corresponds to the state closest to the reference level that also includes the meson flavor-singlet mixing. These four extracted states remain completely consistent after including the mixing effects. The only change worth noting is a slight increase of the errors in some of them. This consistency is not trivial. While the reference states were the ground and first excited states of the separate GEVPs, the ones that include the mixing dynamics are ground state and first excited state for the light meson states but for the charmonium states they are excitations higher up in the ladder. 

\begin{figure}
\includegraphics[width=.95\linewidth]{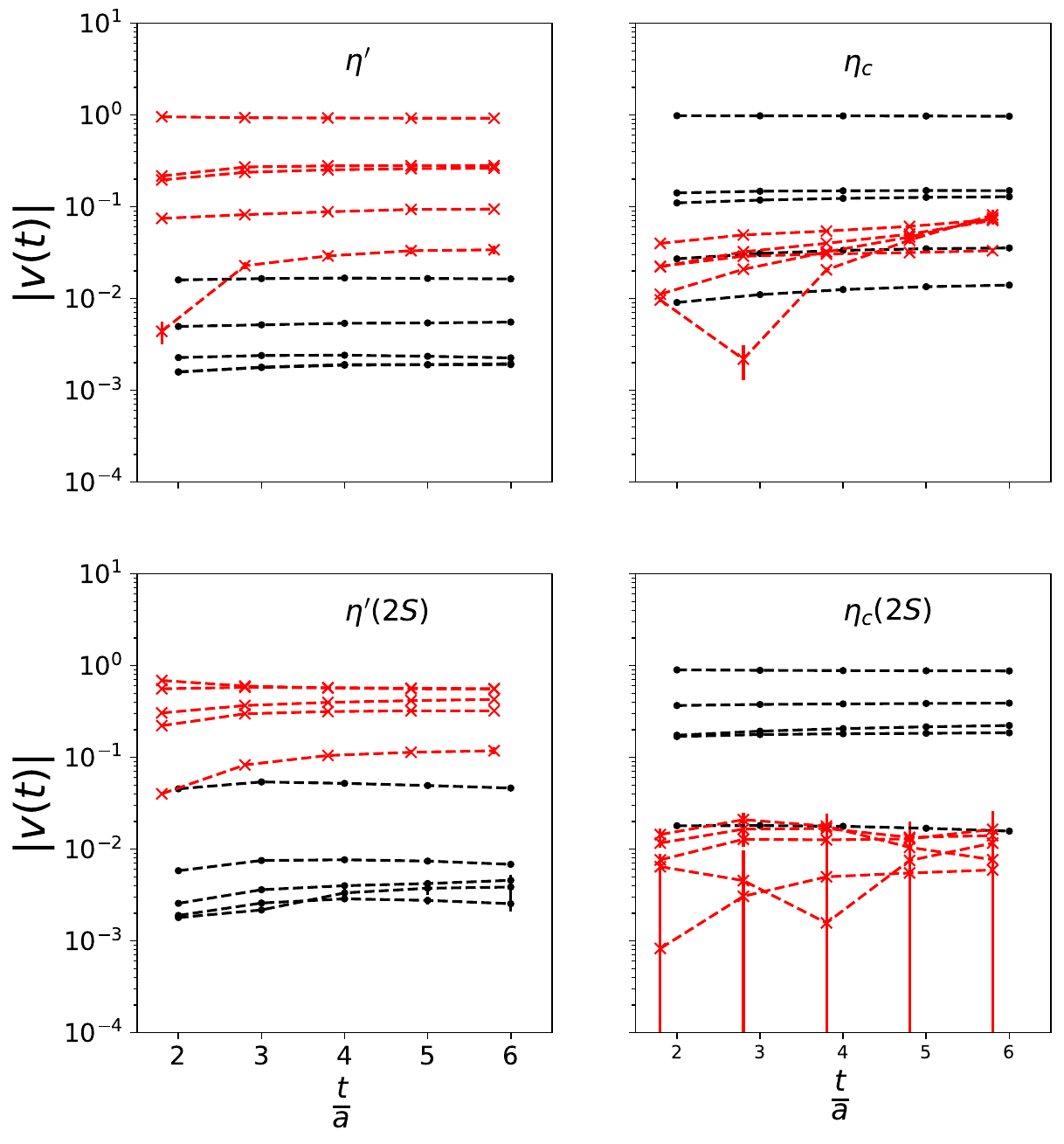}
\caption{Absolute value of the entries of the GEVP vectors corresponding to the four pseudoscalar states of interest in ensemble A1. Black dots correspond to charmonium operators, red crosses to light meson operators and we include the dashed lines connecting the points to guide the eye. }\label{fig:GEVPVectors_G4G5_A1}
\end{figure}

\begin{figure}
\includegraphics[width=.95\linewidth]{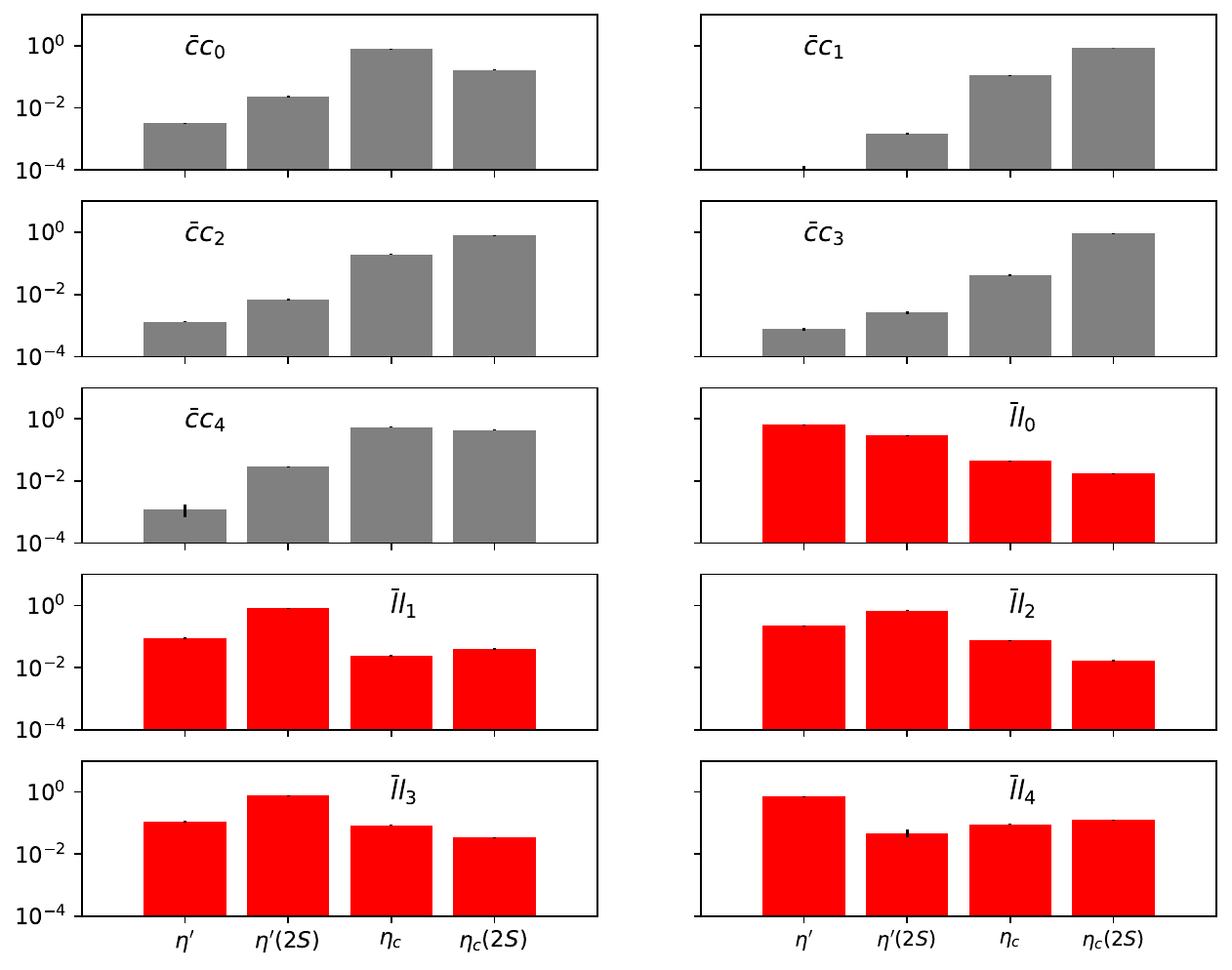}
\caption{Normalized overlaps between the pseudoscalar energy eigenstates of interest and the states created by the light meson and charmonium operators in ensemble A1. Each panel corresponds to a different charmonium or light meson operator, and each bar within a panel corresponds to a different state.}\label{fig:Overlaps_A1mp_A1}
\end{figure}

\begin{figure}
\includegraphics[width=.95\linewidth]{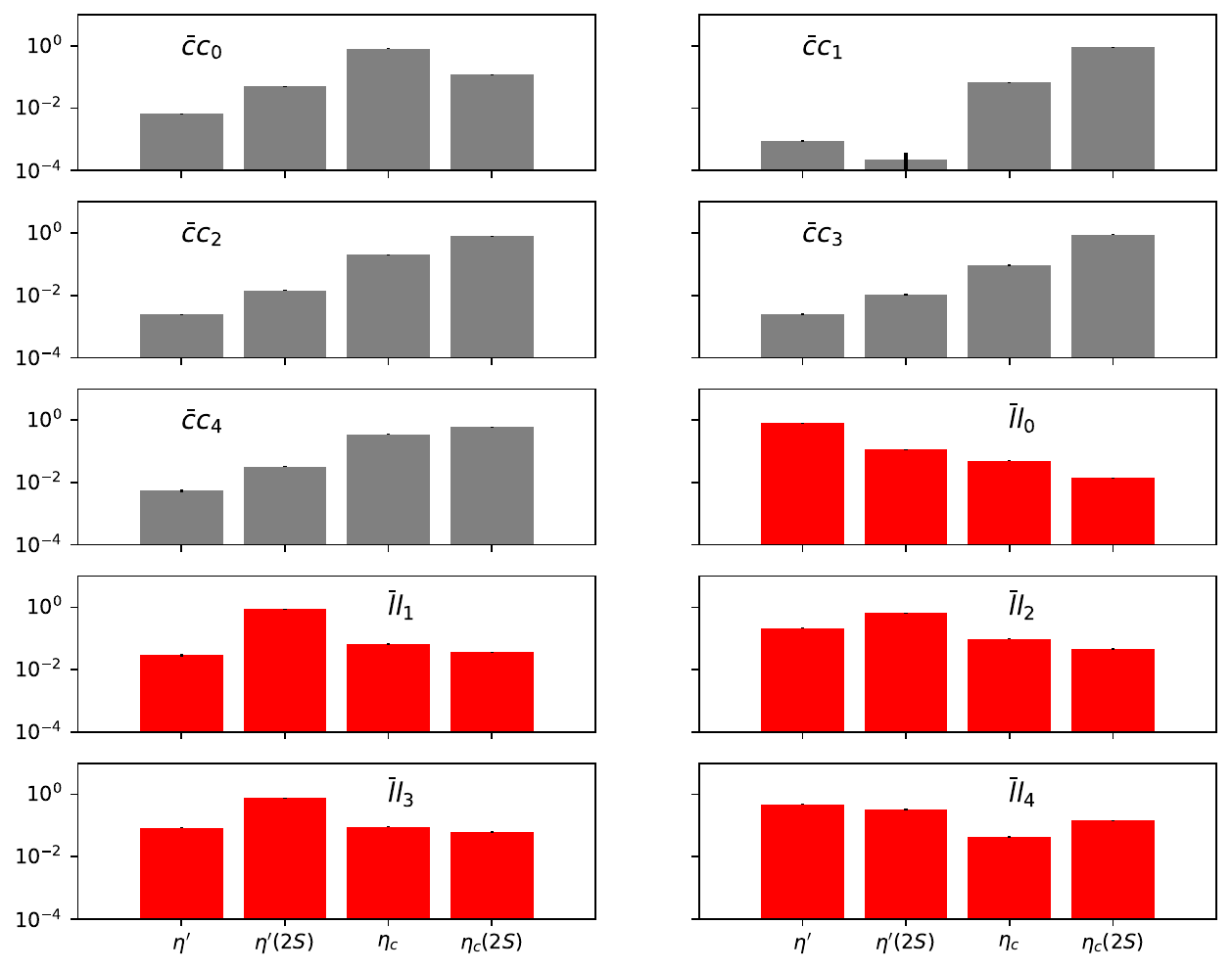}
\caption{Normalized overlaps between the pseudoscalar energy eigenstates of interest and the states created by the light meson and charmonium operators in ensemble A1-heavy.}\label{fig:Overlaps_A1mp_A1h}
\end{figure}

We can get further information about the mixing between light meson and charmonium operators by looking at the GEVP vectors coming from the mixing matrix. Fig.~\ref{fig:GEVPVectors_G4G5_A1} shows the absolute value of the entries of the GEVP vectors corresponding to the different states of interest for ensemble A1. Black dots correspond to charmonium operators, red crosses to light meson operators and we include the dashed lines connecting the points to guide the eye. There are two features of interest in these plots. First, the light meson states $\eta^{\prime}$ and $\eta^{\prime}(2S)$ receive the largest contributions from light meson operators. Similarly, charmonium states $\eta_c$ and $\eta_c(2S)$ receive the largest contributions from charmonium operators. Second, light meson (charmonium) states receive statistically non-zero contributions from charmonium (light meson) operators. From the GEVP vectors we can extract the normalized overlaps mentioned in Sec. \ref{subsec:MixingOverlaps}. For this we use the vectors at $t=2a$. Figs. \ref{fig:Overlaps_A1mp_A1} and \ref{fig:Overlaps_A1mp_A1h} show the normalized overlaps between each of our operators with the four energy eigenstates of interest. As already hinted at by the entries of the GEVP vectors, light meson (charmonium) states mostly overlap with the states created by the light meson (charmonium) operators. Furthermore, mixing overlaps, e.g between charmonium energy eigenstates and states created by light meson operators, are statistically non-zero in some cases. The overall qualitative behavior remains similar over the two pion masses. Interestingly, states created by light meson operators overlap with charmonium energy eigenstates more than states created by charmonium operators overlap with light meson energy eigenstates. This explains why the GEVP which only includes charmonium operators does not yield any states in the light meson region. This is the opposite to what we observed in the $0^{++}$ case in a recent study \cite{urreanino2025flavormixingcharmoniumlight}, where the charmonium operators alone in a GEVP already saw a state in the light meson energy region. 

\begin{figure}
\includegraphics[width=.95\linewidth]{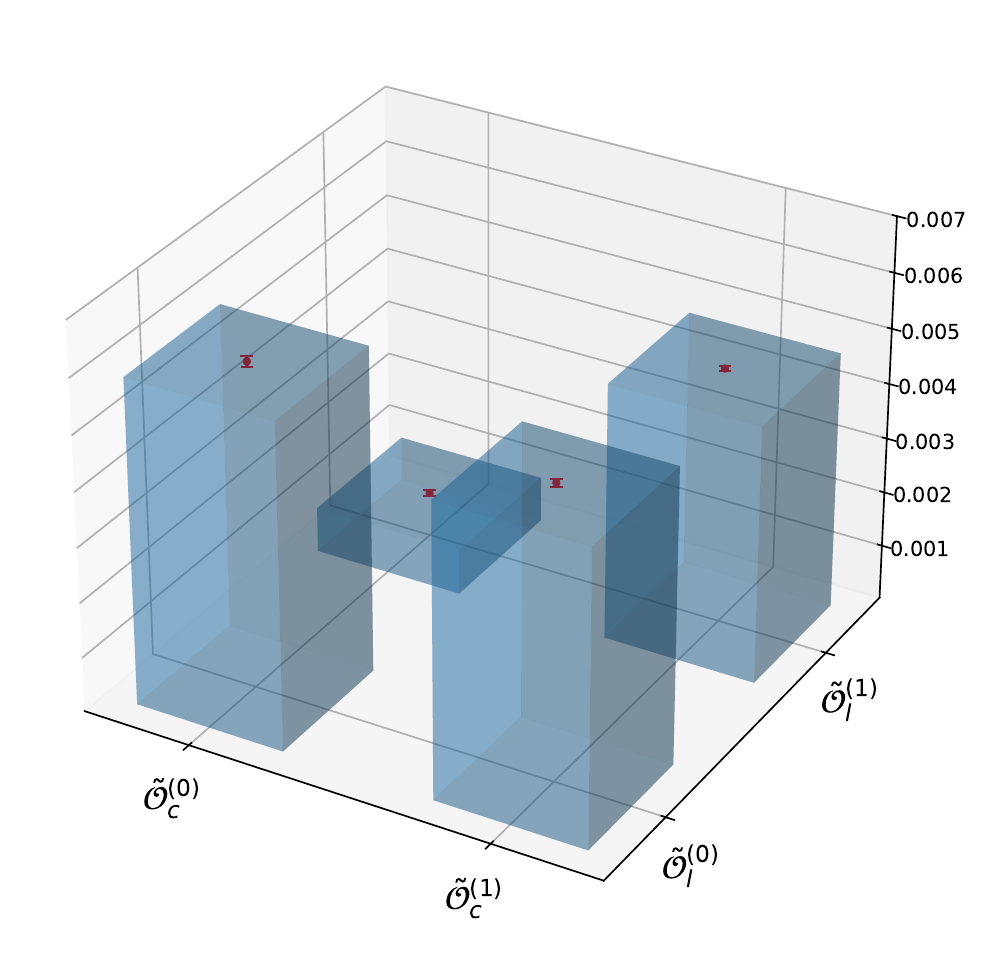}
\caption{Off-diagonal block of the partially diagonalized correlation matrix at $t=a$ for the pseudoscalar channel in ensemble A1. We show the absolute value of the entries.}\label{fig:NewBasis_G4G5_A1}
\end{figure}

\begin{figure}
\includegraphics[width=.95\linewidth]{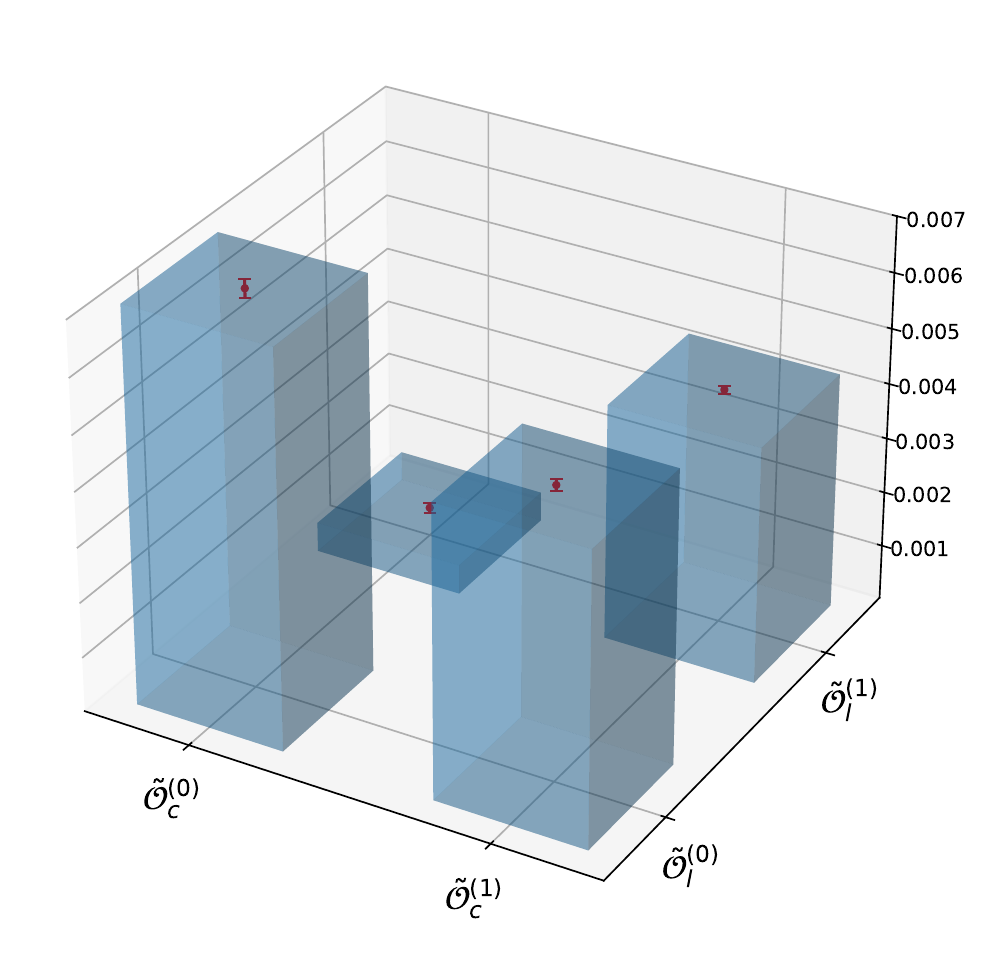}
\caption{Off-diagonal block of the partially diagonalized correlation matrix at $t=a$ for the pseudoscalar channel in ensemble A1-heavy. We show the absolute value of the entries.}\label{fig:NewBasis_G4G5_A1h}
\end{figure}

As a final step, we define an additional correlation matrix, denoted as the "partially diagonalized" correlation matrix, by replacing $V_c$ $\left( V_l  \right)$ in Eq. \ref{eq:PartialPruning} with the vectors of the ground and first excited states which arise from solving a smaller GEVP with $C_{cc}(t)$ $\left( C_{ll}(t)  \right)$. The $2\times 2$ diagonal blocks of this partially diagonalized matrix are the identity matrix at $t = t_G$. Since we are now in a basis where the operators best approximate charmonium and light meson states without any mixing, the remaining $2\times 2$ off-diagonal blocks tell us how much these optimal "pure" charmonium/light meson operators mix with each other. Figs. \ref{fig:NewBasis_G4G5_A1} and \ref{fig:NewBasis_G4G5_A1h} show the absolute value of the $2\times 2$ off-diagonal block. The labels $\Tilde{\mathcal{O}}_{l/c}^{(i)}$ denote the ground (0) and first excited state (1) operators obtained from the smaller GEVPs without mixing at $t=2a$ as described in Sec. \ref{sec:methods}. While the entries of this off-diagonal block are much smaller than the unity diagonal, any vector which fully diagonalizes this matrix needs to have contributions corresponding to both light meson and charmonium operators. This is fully consistent with the GEVP vector entries plotted in Fig. \ref{fig:GEVPVectors_G4G5_A1} and emphasizes how all types of operators sample common energy eigenstates. While the mixing overlaps are considerably small, their effects could become more visible if we manage to reduce statistical uncertainty and reach larger temporal separations. In this case, effective masses from charmonium-only correlation functions would eventually go down to the ground light-meson state.

\subsection{Vector channel}

\begin{figure}
\includegraphics[width=.95\linewidth]{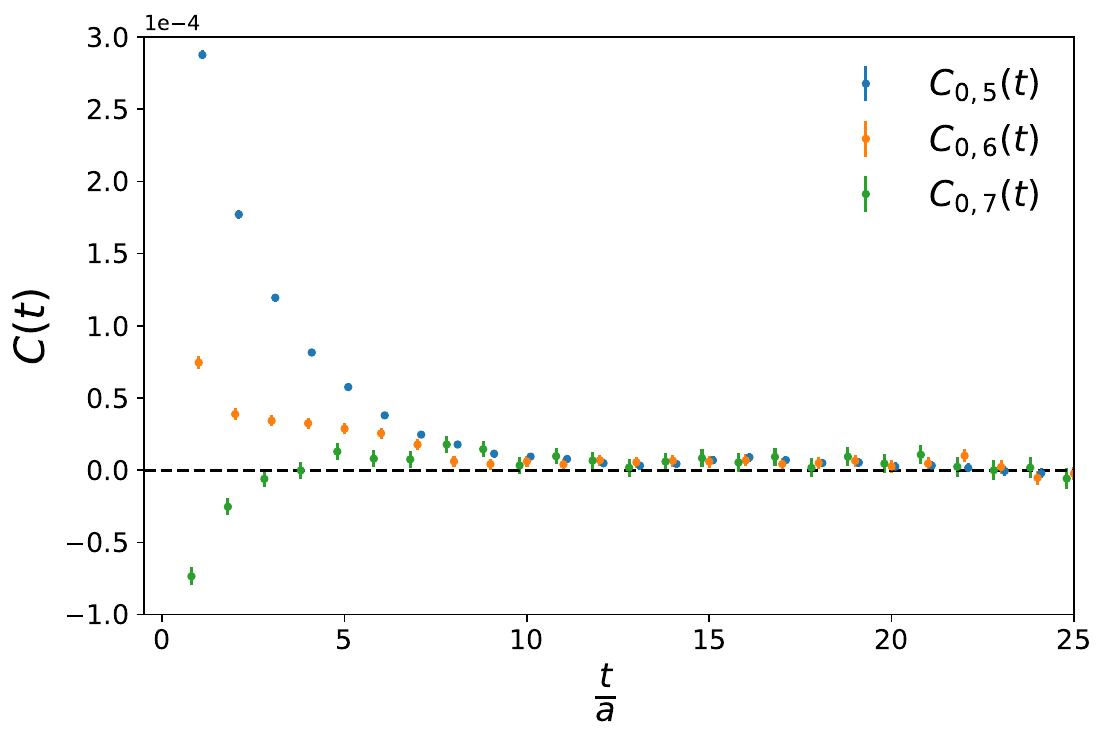}
\caption{Selected off-diagonal vector correlations between the light meson and charmonium operators in ensemble A1.}\label{fig:OffDiagonalCorrelations_Gi_A1}
\end{figure}

\begin{figure}
\includegraphics[width=.95\linewidth]{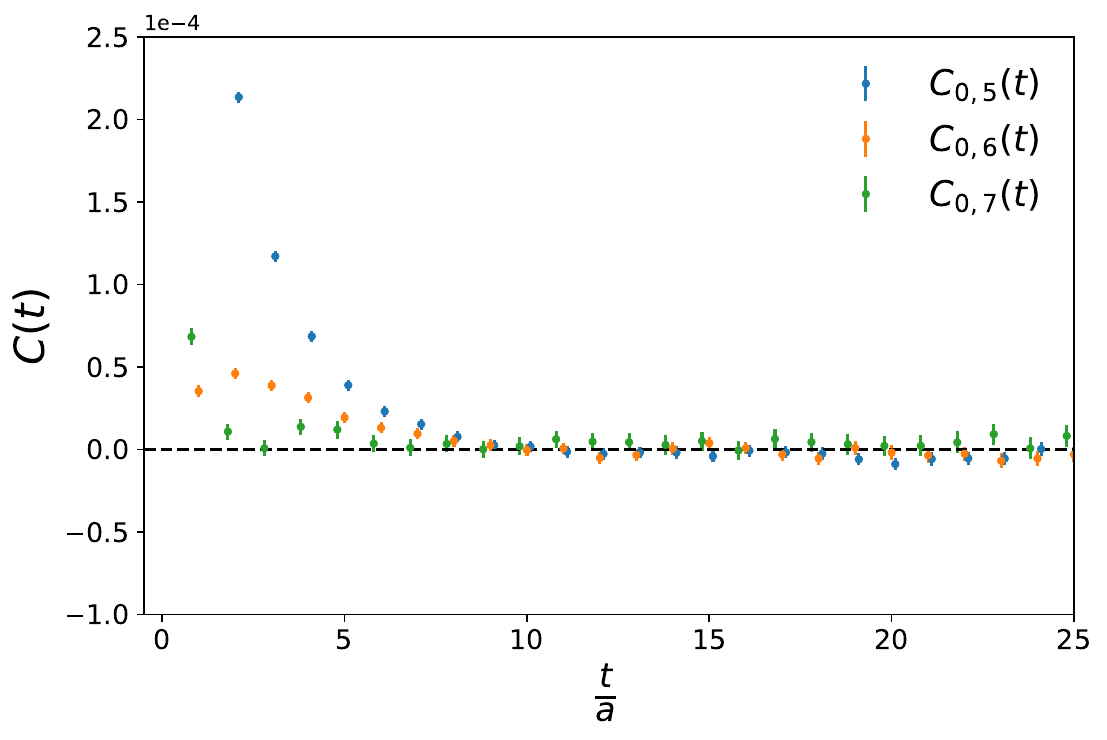}
\caption{Selected off-diagonal vector correlations between the light meson and charmonium operators in ensemble A1-heavy.}\label{fig:OffDiagonalCorrelations_Gi_A1h}
\end{figure}

\begin{figure}
\includegraphics[width=.95\linewidth]{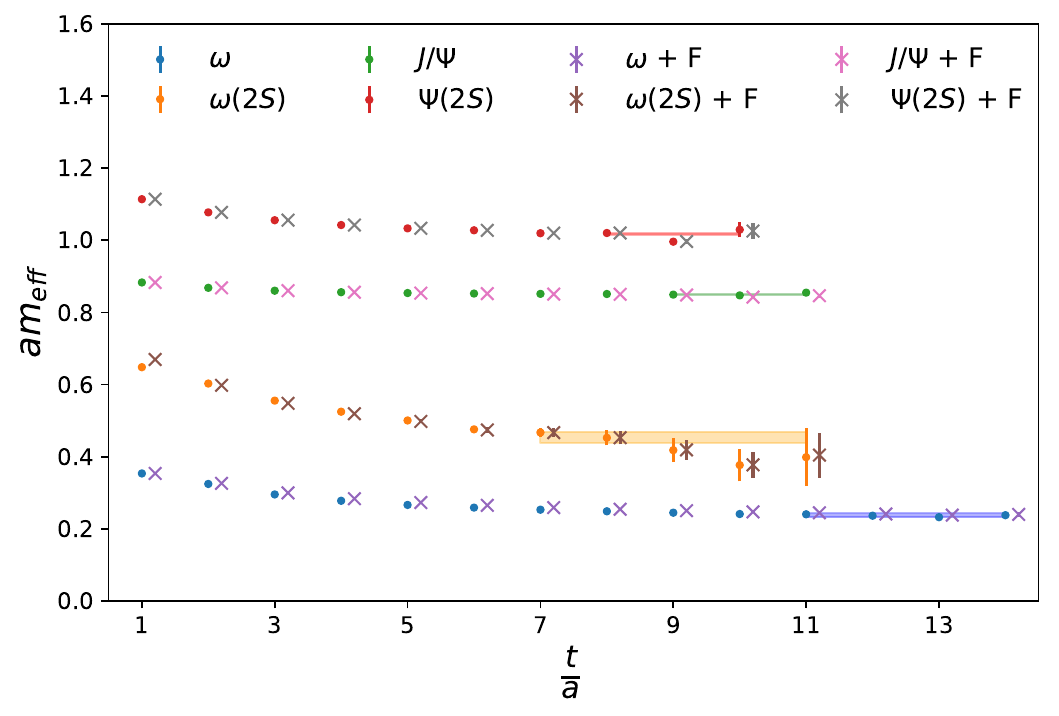}
\caption{Effective masses for the two lightest vector light meson and charmonium states calculated with and without taking into account the mixing between both types of operators in ensemble A1.}\label{fig:T1mm_LightCharmGi_A1}
\end{figure}

\begin{figure}
\includegraphics[width=.95\linewidth]{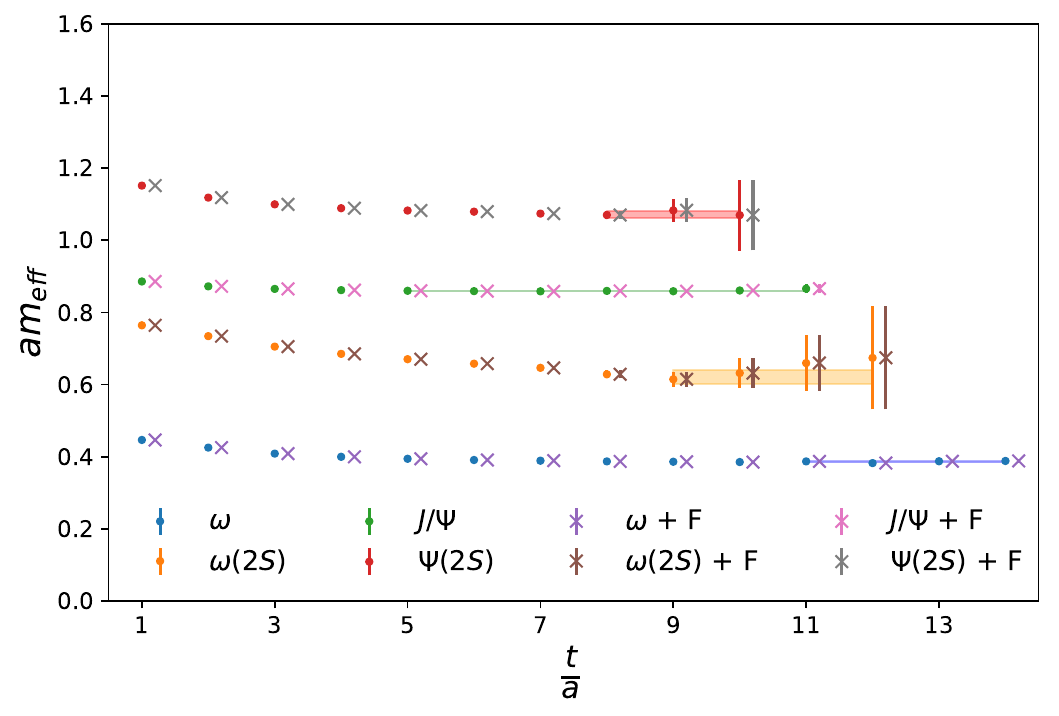}
\caption{Effective masses for the two lightest vector light meson and charmonium states calculated with and without taking into account the mixing between both types of operators in ensemble A1-heavy.}\label{fig:T1mm_LightCharmGi_A1h}
\end{figure}

As with the pseudo-scalar channel, mixing off-diagonal correlations are statistically non-zero at early times. Figs. \ref{fig:OffDiagonalCorrelations_Gi_A1} and \ref{fig:OffDiagonalCorrelations_Gi_A1h} show them for the two ensembles. Again, there are sign changes caused by relative minus signs between terms in the corresponding spectral decomposition. We extract the effective masses for two light meson and charmonium energy eigenstates which we label via separate GEVPs. Figs. \ref{fig:T1mm_LightCharmGi_A1} and \ref{fig:T1mm_LightCharmGi_A1h} show these effective masses obtained with and without taking into account the mixing between the different types of operators. Again, the GEVP involving only charmonium operators does not yield a state in the energy region of the light states. Furthermore, the four states of interest remain fully consistent after introducing the flavor singlet mixing effects. \\



\begin{figure}
\includegraphics[width=.95\linewidth]{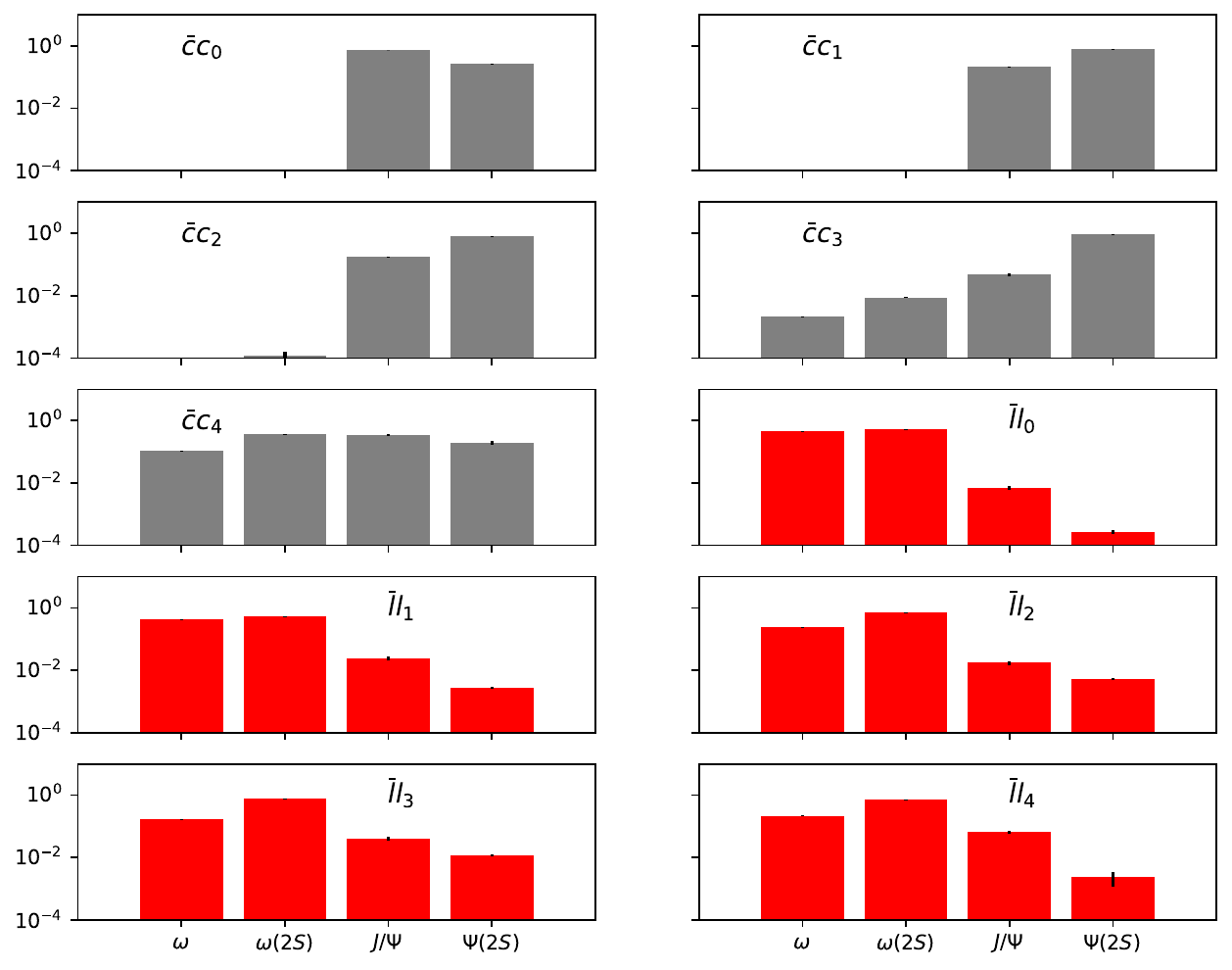}
\caption{Normalized overlaps between the vector energy eigenstates of interest and the states created by the light meson and charmonium operators in ensemble A1.}\label{fig:Overlaps_T1mm_A1}
\end{figure}

\begin{figure}
\includegraphics[width=.95\linewidth]{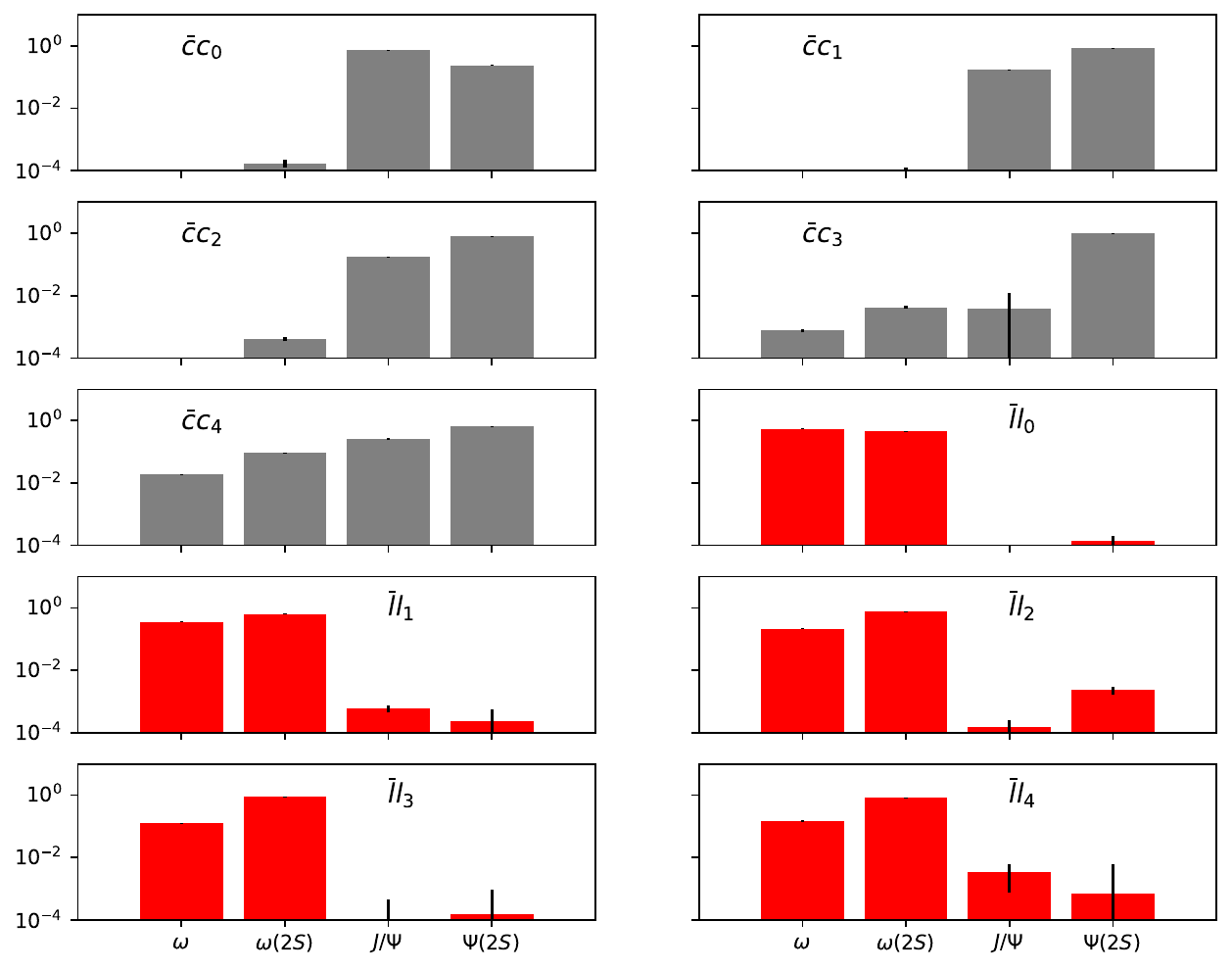}
\caption{Normalized overlaps between the vector energy eigenstates of interest and the states created by the light meson and charmonium operators in ensemble A1-heavy.}\label{fig:Overlaps_T1mm_A1h}
\end{figure}

\begin{figure}
\includegraphics[width=.95\linewidth]{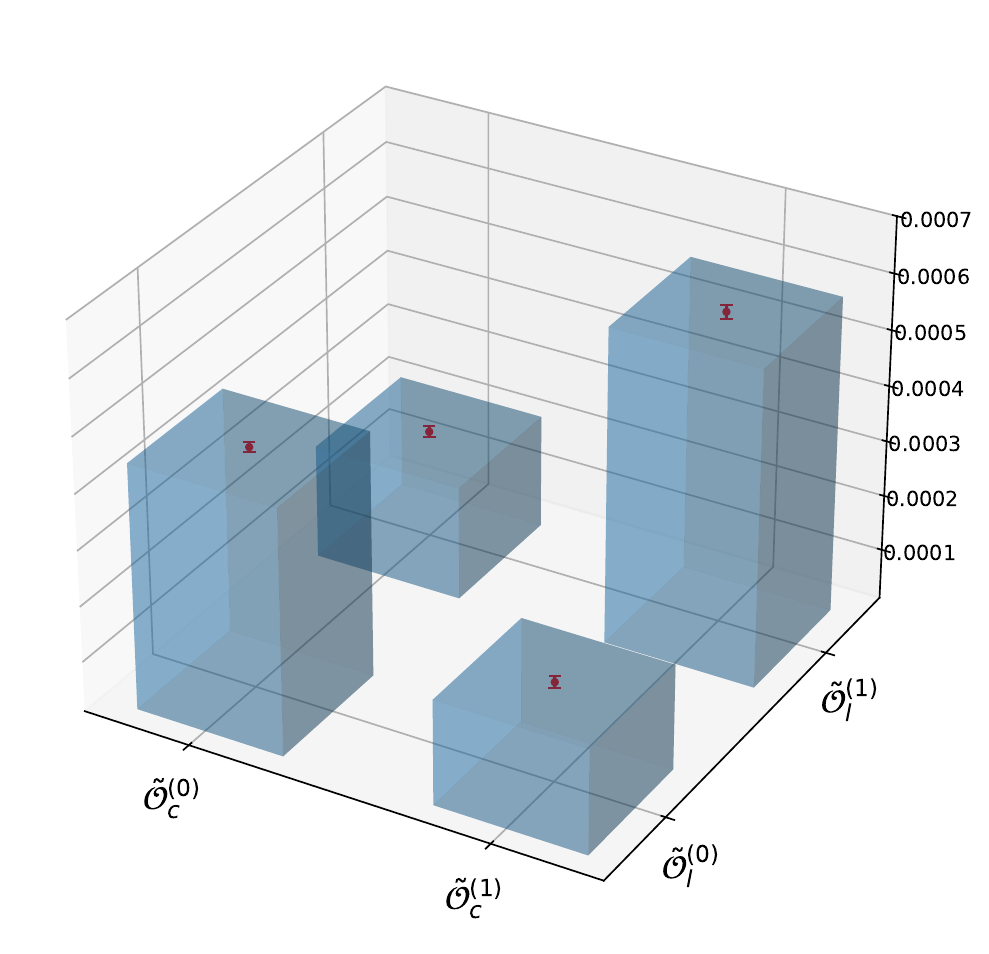}
\caption{Off-diagonal block of the partially diagonalized correlation matrix at $t=a$ for the vector channel in ensemble A1. We show the absolute value of the entries.}\label{fig:NewBasis_Gi_A1}
\end{figure}

\begin{figure}[H]
\includegraphics[width=.95\linewidth]{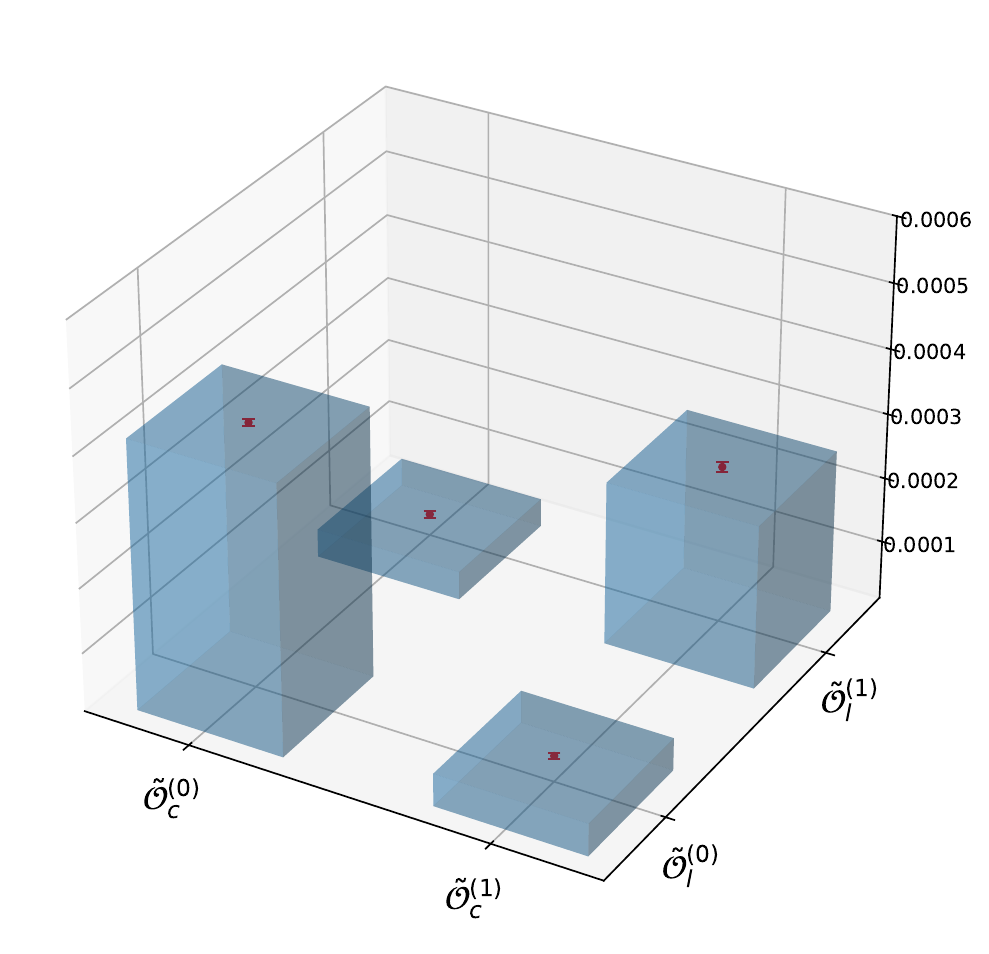}
\caption{Off-diagonal block of the partially diagonalized correlation matrix at $t=a$ for the vector channel in ensemble A1-heavy. We show the logarithm of the absolute value of the entries.}\label{fig:NewBasis_Gi_A1h}
\end{figure}

To quantify the mixing we repeat the analysis of the previous section by checking the mixing overlaps which are displayed in Figs. \ref{fig:Overlaps_T1mm_A1} and \ref{fig:Overlaps_T1mm_A1h}, where the qualitative behavior is not affected much by the pion mass difference as for the pseudoscalar case. Except for the fourth and fifth charmonium operator, all other operators create states which almost exclusively overlap onto their corresponding type of energy eigenstate, i.e light meson operators operators overlap mainly with light meson states, and charmonium operators do so with charmonium states. The overlaps of the fifth charmonium operator hint that at sufficiently large times the corresponding effective masses should see one of the light meson states, unfortunately the corresponding correlation function is one of the noisiest and the signal is lost before any light state can be resolved. Finally, in Figs. \ref{fig:NewBasis_Gi_A1} and \ref{fig:NewBasis_Gi_A1h} we show the $2\times 2$ off-diagonal block of the partially diagonalized matrix for the vector channel. As for the pseudoscalar channel, the off-diagonal blocks are small yet non-zero. Furthermore, we find them to be significantly smaller than in the pseudoscalar case, which is consistent with the OZI suppression due to three gluon exchange. 

\subsection{Towards mixing including gluonic operators}

\begin{figure}
\includegraphics[width=.95\linewidth]{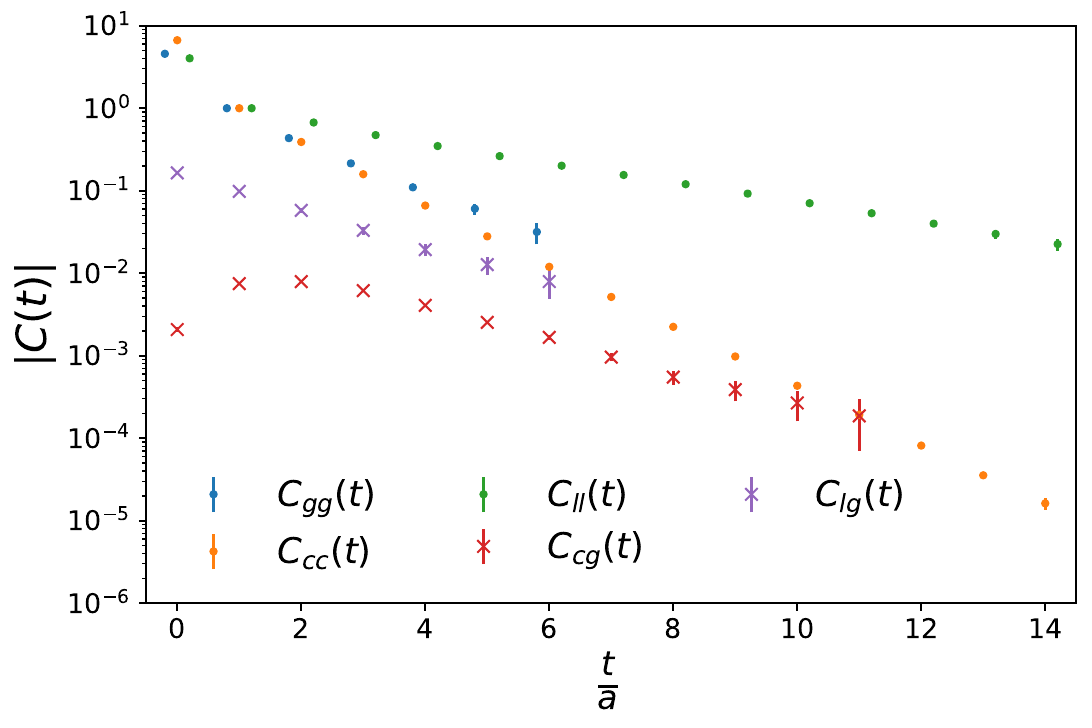}
\caption{Absolute value of selected entries of the pseudoscalar correlation matrix involving light meson, charmonium and Wilson loop operators in ensemble A1.}\label{fig:WithLoops_A1mp_A1}
\end{figure}

Light meson and charmonium are not the only flavor-singlet one-particle operators which can be included in the mixing matrix. Spatial Wilson loops such as the ones used in \cite{Berg, Barca2024, Morningstar:1999rf} define operators with the same quantum numbers. While extending the basis with a different type of operators is a natural way to better resolve the energy spectrum, such gluonic operators are known to have a severe signal-to-noise problem in their temporal correlation functions and therefore require significantly more statistics to attain a quality comparable to meson correlations. While multi-level integration has proven successful in tackling this problem in pure-gauge theory, e.g as in \cite{Barca2024}, there is still work to be done to use this technique with dynamical quarks. Furthermore, previous studies of Wilson loop operators in the presence of dynamical quarks \cite{Athenodorou2023} as well as in pure-gauge theory \cite{Morningstar:1999rf} place the pseudoscalar glueball mass above the scalar and tensor one, meaning we need to resolve a somewhat heavy state at $\approx 2.6$ GeV. The quenched prediction \cite{Morningstar:1999rf} of the vector glueball at $\approx 3.9$ GeV puts it above our $\Psi (2S)$ state, making it less relevant for this work. Since we need large statistics to get a first signal of the Wilson loop correlations and the mixing correlations with the light meson and charmonium operators, we only use the A1 ensemble where we currently have 9000 configurations. Fig. \ref{fig:WithLoops_A1mp_A1} shows the absolute value of selected entries of a correlation matrix which now involves the mixing between the meson operators and the gluonic ones built from APE-smeared \cite{Albanese:1987ds} Wilson loops with shapes taken from \cite{Barca2024}. Here $C_{gg}(t)$ is the diagonal correlation of a Wilson loop operator and $C_{ig}(t)$, $i=l,c$, are the mixing correlations between the Wilson loop operator and a light meson or charmonium operator respectively. We observe statistically non-zero signal for all of these correlations and is a first step towards extending our systematic study to include the mixing with these gluonic operators. The main problem is how quickly $C_{gg}(t)$ loses significance. Since all plateaus calculated in this work start around $t=6a$, we need to at least double the number of meaningful data points before having a sufficiently precise mixing correlation matrix to solve a GEVP. Nonetheless, the presence of a small but non-zero signal for the mixing correlation $C_{cg}(t)$ is consistent with previous studies which have also measured these mixing correlations albeit only with dynamical charm quarks. Given the proposal of $X(2370)$ as a pseudoscalar glueball, a study of how gluonic operators mix with mesonic ones as a extension of this work becomes very relevant.

\section{Mass splittings}

\begin{figure}
\includegraphics[width=.95\linewidth]{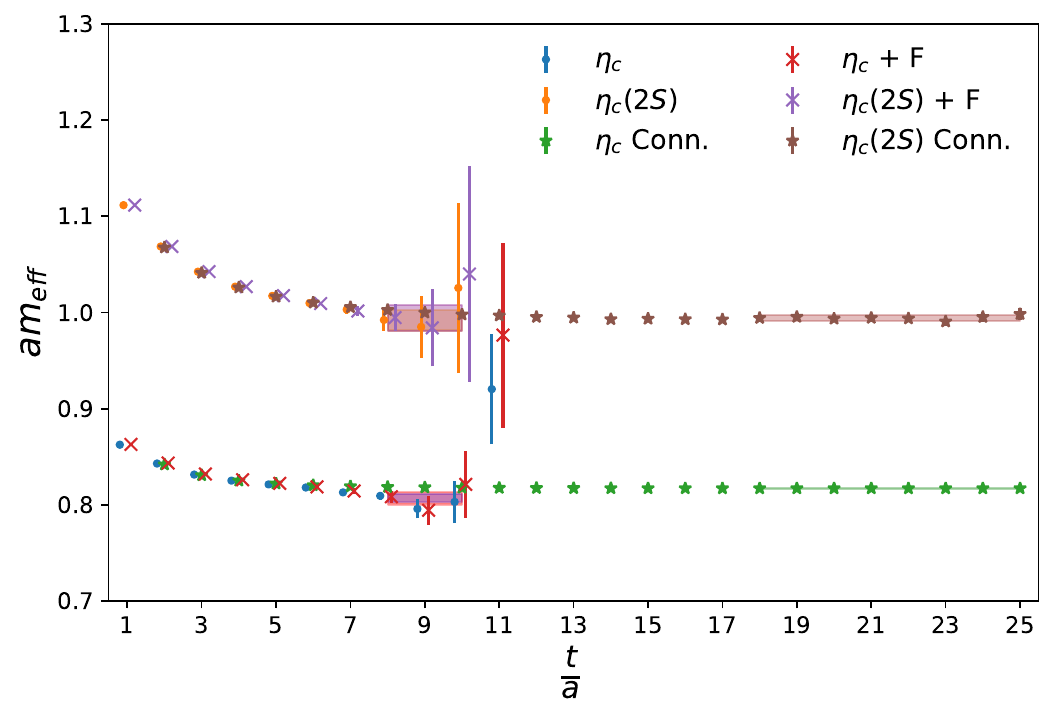}
\caption{Effective masses of the $\eta_c$ and $\eta_c(2S)$ states using only the quark-connected correlations, quark-connected plus quark-disconnected correlations and with explicit flavor-singlet mixing in ensemble A1.}\label{fig:A1mp_ConnVsFull_A1}
\end{figure}

\begin{figure}
\includegraphics[width=.95\linewidth]{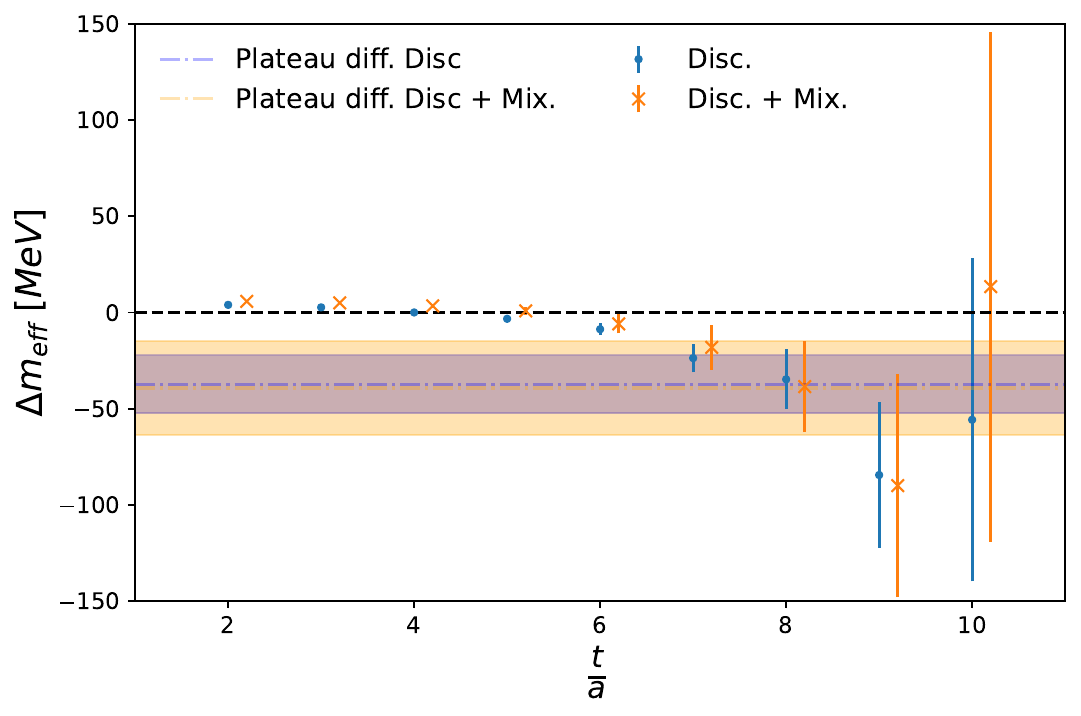}
\caption{Effective mass splitting $\Delta m_{\text{eff.}} = m_{\eta_c} - m_{\eta_c\ \text{Conn.}}$ with and without the effects of explicit flavor-singlet mixing in ensemble A1. The bands correspond to the difference between the plateaus shown in Fig. \ref{fig:A1mp_ConnVsFull_A1}.}
\label{fig:Splitting_A1mp_A1}
\end{figure}

Mass splittings are useful benchmark quantities to quantify the effects of the disconnected correlations computed in this work. These effects are often ignored based on the expected OZI suppression of these disconnected contributions. Nonetheless, directly measuring them is important since with a single charm quark, as in nature, only the flavor-singlet charmonium correlations are physically meaningful, i.e charm-connected correlations alone are not in a well-defined symmetry channel. Fig.~\ref{fig:A1mp_ConnVsFull_A1} shows the effective masses for the $\eta_c$ and $\eta_c(2S)$ with and without the effects of disconnected and mixing correlations. Once these effects are included, the errors increase significantly and the signal is lost at early time separations. We cannot reach the time separations where the mass plateau is extracted for the connected-only correlations, nonetheless we attempt a plateau calculation with the available points. While for the $\eta_c(2S)$ the plateau values remain consistent with and without the disconnected and mixing effects, the $\eta_c$ mass has a slight shift downwards of $-37(15)$ MeV with the charm-disconnected effects and $-39(24)$ MeV with the mixing effects. A negative mass shift with a magnitude of $20 - 30$ MeV had been measured in the past \cite{Forcrand, McNeile, McNeil2} and a small shift of $\approx -3.1$ MeV is predicted by NRQCD perturbation theory \cite{Follana, Braaten} yet it conflicts with recent direct and indirect lattice calculations of a small positive shift of $\approx 7.3(1.2)$ MeV \cite{Hatton, Zhang_2022}, although the direct calculation considers 2 degenerate charm quarks while we work with only one. As an alternative calculation of this mass shift, we look at the ratio
\begin{align}
    R(t) &= \frac{C_{\text{Full}}(t)}{C_{\text{Conn.}}(t)},
    \label{eqn:Ratio}
\end{align}
where "Full" means either with disconnected correlations or with disconnected and mixing correlations. Previous lattice studies extracted the shift from the slope of $R(t) - 1$ \cite{Forcrand, McNeile, McNeil2}. We use the model 
\begin{align}
    \lim_{t\rightarrow\infty} R(t) = Ae^{-\Delta m t},
    \label{eqn:FitRatio}
\end{align}
with $\Delta m = m_{\eta_c\ \text{Full}} - m_{\eta_c\ \text{Conn.}}$ to analyse this ratio. If the connected-only correlation $C_{\text{Conn.}}(t)$ had a spectral decomposition then Eqn.~\ref{eqn:FitRatio} would hold in the limit of large $t$. This is not the case in our setup however, nonetheless it is not a bad approximation since, as we saw for the scale setting procedure, the connected-only correlations exhibit an exponential-like behavior suitable for extracting effective masses. Fig.~\ref{fig:Splitting_A1mp_A1} shows the effective mass determination of $\Delta m$. As a reference, we also include the horizontal bands showing the difference between the mass plateau with disconnected (and mixing) and the connected-only mass plateau. Both determinations of the shift are in agreement in terms of sign at time separations larger than $t = 6a$. At very early time separations the effective mass splitting is positive and very small; $3.95(11)$ MeV with disconnected and $5.75(19)$ MeV with disconnected and mixing effects. These values agree in sign and magnitude with the recent direct and indirect lattice determinations yet we observe the change of sign and magnitude as time separation increases.\\

The hyperfine splitting is another useful quantity to test the quality of our results since we can compare it directly with the value from experiment. We define the relative hyperfine splitting for ground and first excited states as
\begin{align}
    \delta_{\text{HF}}^{0} &= \frac{m_{J/\psi} - m_{\eta_c}}{m_{J/\psi}} 
           \nonumber\\
    \delta_{\text{HF}}^{1} &= \frac{m_{\psi (2S)} - m_{\eta_c(2S)}}{m_{\psi (2S)}},
    \label{eq:deltaHF}
\end{align}
where we divide by the vector mass in both cases to avoid the need to convert to physical units as well as keeping the effects of the disconnected contributions only in the numerator.We include disconnected and mixing effects only for the pseudoscalar particles, where they are more significant. We summarize our results in Table \ref{table:HF_A1}. The results from connected-only correlations are in good agreement with experiment as well as with other lattice studies \cite{Hollwieser:2020qri, DeTar:2018uko}. The upward shift in the ground state splitting is due to the downward shift of the $\eta_c$ mass we observed. The deviation from the connected-only value is around $2.5 \sigma$. This is not unexpected, since we observe a mass shift of roughly $37$ MeV and the hyperfine splitting in nature is around $113$ MeV. A more precise determination of this shift could help better pinpoint its effects on this splitting. 

\begin{center}
\begin{table}
\begin{tabular}{||c c c c c||} 
 \hline
 Splitting & Conn. & + Disc. & + Disc. $\&$ Mix. & PDG \cite{ParticleDataGroup:2024cfk}\\ [0.5ex] 
 \hline\hline
 $\delta_{\text{HF}}^{0}$ & 0.03671(10) & 0.0483(47) & 0.0489(76) & 0.03642\\ 
 \hline
 $\delta_{\text{HF}}^{1}$ & 0.0163(13) & 0.019(11) & 0.016(14) & 0.0131\\ [1ex] 
 \hline
\end{tabular}
\caption{Normalised ground and first excited state hyperfine splittings in 
  ensemble A1 (Eqn.~\ref{eq:deltaHF}).}
\label{table:HF_A1}
\end{table}
\end{center}

\section{Conclusions}

In this work we presented a systematic study of implicit and explicit mixing between flavor-singlet charmonium and light meson pseudoscalar and vector operators at two different pion masses. For each symmetry channel we built a correlation matrix based on both types of flavor-singlet operators using distillation profiles and obtained statistically significant signal for the off-diagonal entries relating light meson and charmonium operators, which are often neglected in the literature based on the OZI suppression argument. We obtained non-zero overlaps between the state created by our light meson and charmonium operators and the energy eigenstates in the light meson and charmonium energy regions. One major consequence of these non-zero overlaps is that effective masses coming from a charmonium operator temporal correlation should eventually plateau at the mass of the ground state in the light meson energy region. To be able to reliably extract effective masses of higher excitations, including those in the charmonium energy region, we solved different GEVPs based on the mixing correlation matrix. As a reference, we solved separate charmonium and light meson GEVPs where explicit mixing between these two types of operators is not taken into account. We then solved GEVPs which include this explicit mixing and compared the resulting spectra in both cases. The effective masses for each state remained consistent with each other and the presence of statistically non-zero explicit mixing does not have a major impact on them. This is not entirely unexpected; we observed that states created by charmonium (light meson) operators overlap mostly with charmonium (light meson) energy eigenstates in the different regions of energy. We also presented a first approach to include spatial Wilson loop operators in the operator basis to sample possible glueball states, but the currently available statistics are still too low for purely gluonic operators and we defer the full mixing including this type of operators to a future study. Since we could access at least the two lightest light meson and charmonium pseudoscalar and vector states we measured the hyperfine splitting in both energy levels both with and without the effects of flavor-singlet mixing. Both approaches yielded results consistent with each other as well as very close to the PDG values. This work complements our ongoing study of the flavor-singlet mixing in the scalar channel, where the effects of flavor-singlet mixing were found to be much more significant in \cite{urreanino2025flavormixingcharmoniumlight} than for the channels considered in this work.

\begin{acknowledgments}
The authors gratefully acknowledge the Gauss Centre for Supercomputing e.V. (www.gauss-centre.eu) for funding this project by providing computing time on the GCS Supercomputer SuperMUC-NG at Leibniz Supercomputing Centre (www.lrz.de) and the scientific support and HPC resources provided by the Erlangen National High Performance Computing Center (NHR@FAU) of the Friedrich-Alexander-Universität Erlangen-Nürnberg (FAU) under the NHR project k103bf. NHR funding is provided by federal and Bavarian state authorities. NHR@FAU hardware is partially funded by the German Research Foundation (DFG) – 440719683. We also thank Juelich Supercomputing Centre (JSC) for allocating space on the storage system JUST (project ID hwu35).
J.A.U.N. acknowledges support from a Science Foundation Ireland (Research Ireland) Frontiers for the Future Project award [grant number SFI-21/FFP-P/10186]. M.P. was supported by the European Union’s Horizon 2020 research and innovation programme under grant agreement 824093 (STRONG-2020). J.F. acknowledges financial support by the Eric \& Wendy Schmidt Fund for Strategic Innovation through the CERN Next Generation Triggers project under grant agreement number SIF-2023-004. The work is supported by the German Research Foundation (DFG) research unit FOR5269 "Future methods for studying confined gluons in QCD". R.H. acknowledges funding from the programme " Netzwerke 2021", an initiative of the Ministry of Culture and Science of the State of Northrhine Westphalia, in the NRW-FAIR network, funding code NW21-024-A. The sole responsibility for the content of this publication lies with the authors. 
\end{acknowledgments}


\bibliography{references}

\begin{thebibliography}{50}%
\makeatletter
\providecommand \@ifxundefined [1]{%
 \@ifx{#1\undefined}
}%
\providecommand \@ifnum [1]{%
 \ifnum #1\expandafter \@firstoftwo
 \else \expandafter \@secondoftwo
 \fi
}%
\providecommand \@ifx [1]{%
 \ifx #1\expandafter \@firstoftwo
 \else \expandafter \@secondoftwo
 \fi
}%
\providecommand \natexlab [1]{#1}%
\providecommand \enquote  [1]{``#1''}%
\providecommand \bibnamefont  [1]{#1}%
\providecommand \bibfnamefont [1]{#1}%
\providecommand \citenamefont [1]{#1}%
\providecommand \href@noop [0]{\@secondoftwo}%
\providecommand \href [0]{\begingroup \@sanitize@url \@href}%
\providecommand \@href[1]{\@@startlink{#1}\@@href}%
\providecommand \@@href[1]{\endgroup#1\@@endlink}%
\providecommand \@sanitize@url [0]{\catcode `\\12\catcode `\$12\catcode `\&12\catcode `\#12\catcode `\^12\catcode `\_12\catcode `\%12\relax}%
\providecommand \@@startlink[1]{}%
\providecommand \@@endlink[0]{}%
\providecommand \url  [0]{\begingroup\@sanitize@url \@url }%
\providecommand \@url [1]{\endgroup\@href {#1}{\urlprefix }}%
\providecommand \urlprefix  [0]{URL }%
\providecommand \Eprint [0]{\href }%
\providecommand \doibase [0]{https://doi.org/}%
\providecommand \selectlanguage [0]{\@gobble}%
\providecommand \bibinfo  [0]{\@secondoftwo}%
\providecommand \bibfield  [0]{\@secondoftwo}%
\providecommand \translation [1]{[#1]}%
\providecommand \BibitemOpen [0]{}%
\providecommand \bibitemStop [0]{}%
\providecommand \bibitemNoStop [0]{.\EOS\space}%
\providecommand \EOS [0]{\spacefactor3000\relax}%
\providecommand \BibitemShut  [1]{\csname bibitem#1\endcsname}%
\let\auto@bib@innerbib\@empty
\bibitem [{\citenamefont {Brambilla}\ \emph {et~al.}(2020)\citenamefont {Brambilla}, \citenamefont {Eidelman}, \citenamefont {Hanhart}, \citenamefont {Nefediev}, \citenamefont {Shen}, \citenamefont {Thomas}, \citenamefont {Vairo},\ and\ \citenamefont {Yuan}}]{Brambilla:2019esw}%
  \BibitemOpen
  \bibfield  {author} {\bibinfo {author} {\bibfnamefont {N.}~\bibnamefont {Brambilla}}, \bibinfo {author} {\bibfnamefont {S.}~\bibnamefont {Eidelman}}, \bibinfo {author} {\bibfnamefont {C.}~\bibnamefont {Hanhart}}, \bibinfo {author} {\bibfnamefont {A.}~\bibnamefont {Nefediev}}, \bibinfo {author} {\bibfnamefont {C.-P.}\ \bibnamefont {Shen}}, \bibinfo {author} {\bibfnamefont {C.~E.}\ \bibnamefont {Thomas}}, \bibinfo {author} {\bibfnamefont {A.}~\bibnamefont {Vairo}},\ and\ \bibinfo {author} {\bibfnamefont {C.-Z.}\ \bibnamefont {Yuan}},\ }\bibfield  {title} {\bibinfo {title} {{The $XYZ$ states: experimental and theoretical status and perspectives}},\ }\href {https://doi.org/10.1016/j.physrep.2020.05.001} {\bibfield  {journal} {\bibinfo  {journal} {Phys. Rept.}\ }\textbf {\bibinfo {volume} {873}},\ \bibinfo {pages} {1} (\bibinfo {year} {2020})},\ \Eprint {https://arxiv.org/abs/1907.07583} {arXiv:1907.07583 [hep-ex]} \BibitemShut {NoStop}%
\bibitem [{\citenamefont {Godfrey}\ and\ \citenamefont {Olsen}(2008)}]{Godfrey:2008nc}%
  \BibitemOpen
  \bibfield  {author} {\bibinfo {author} {\bibfnamefont {S.}~\bibnamefont {Godfrey}}\ and\ \bibinfo {author} {\bibfnamefont {S.~L.}\ \bibnamefont {Olsen}},\ }\bibfield  {title} {\bibinfo {title} {{The Exotic XYZ Charmonium-like Mesons}},\ }\href {https://doi.org/10.1146/annurev.nucl.58.110707.171145} {\bibfield  {journal} {\bibinfo  {journal} {Ann. Rev. Nucl. Part. Sci.}\ }\textbf {\bibinfo {volume} {58}},\ \bibinfo {pages} {51} (\bibinfo {year} {2008})},\ \Eprint {https://arxiv.org/abs/0801.3867} {arXiv:0801.3867 [hep-ph]} \BibitemShut {NoStop}%
\bibitem [{\citenamefont {Ablikim}\ \emph {et~al.}(2024)\citenamefont {Ablikim} \emph {et~al.}}]{AblikimGlueball}%
  \BibitemOpen
  \bibfield  {author} {\bibinfo {author} {\bibfnamefont {M.}~\bibnamefont {Ablikim}} \emph {et~al.} (\bibinfo {collaboration} {BESIII Collaboration}),\ }\bibfield  {title} {\bibinfo {title} {Determination of spin-parity quantum numbers of $x(2370)$ as ${0}^{\ensuremath{-}+}$ from $j/\ensuremath{\psi}\ensuremath{\rightarrow}\ensuremath{\gamma}{K}_{S}^{0}{K}_{S}^{0}{\ensuremath{\eta}}^{\ensuremath{'}}$},\ }\href {https://doi.org/10.1103/PhysRevLett.132.181901} {\bibfield  {journal} {\bibinfo  {journal} {Phys. Rev. Lett.}\ }\textbf {\bibinfo {volume} {132}},\ \bibinfo {pages} {181901} (\bibinfo {year} {2024})}\BibitemShut {NoStop}%
\bibitem [{\citenamefont {Liu}\ \emph {et~al.}(2012)\citenamefont {Liu}, \citenamefont {Moir}, \citenamefont {Peardon}, \citenamefont {Ryan}, \citenamefont {Thomas}, \citenamefont {Vilaseca}, \citenamefont {Dudek}, \citenamefont {Edwards}, \citenamefont {Joo},\ and\ \citenamefont {Richards}}]{Liu:2012ze}%
  \BibitemOpen
  \bibfield  {author} {\bibinfo {author} {\bibfnamefont {L.}~\bibnamefont {Liu}}, \bibinfo {author} {\bibfnamefont {G.}~\bibnamefont {Moir}}, \bibinfo {author} {\bibfnamefont {M.}~\bibnamefont {Peardon}}, \bibinfo {author} {\bibfnamefont {S.~M.}\ \bibnamefont {Ryan}}, \bibinfo {author} {\bibfnamefont {C.~E.}\ \bibnamefont {Thomas}}, \bibinfo {author} {\bibfnamefont {P.}~\bibnamefont {Vilaseca}}, \bibinfo {author} {\bibfnamefont {J.~J.}\ \bibnamefont {Dudek}}, \bibinfo {author} {\bibfnamefont {R.~G.}\ \bibnamefont {Edwards}}, \bibinfo {author} {\bibfnamefont {B.}~\bibnamefont {Joo}},\ and\ \bibinfo {author} {\bibfnamefont {D.~G.}\ \bibnamefont {Richards}} (\bibinfo {collaboration} {Hadron Spectrum}),\ }\bibfield  {title} {\bibinfo {title} {{Excited and exotic charmonium spectroscopy from lattice QCD}},\ }\href {https://doi.org/10.1007/JHEP07(2012)126} {\bibfield  {journal} {\bibinfo  {journal} {JHEP}\ }\textbf {\bibinfo {volume} {07}},\ \bibinfo {pages} {126}},\ \Eprint {https://arxiv.org/abs/1204.5425}
  {arXiv:1204.5425 [hep-ph]} \BibitemShut {NoStop}%
\bibitem [{\citenamefont {Cheung}\ \emph {et~al.}(2016)\citenamefont {Cheung}, \citenamefont {O'Hara}, \citenamefont {Moir}, \citenamefont {Peardon}, \citenamefont {Ryan}, \citenamefont {Thomas},\ and\ \citenamefont {Tims}}]{Cheung:2016bym}%
  \BibitemOpen
  \bibfield  {author} {\bibinfo {author} {\bibfnamefont {G.~K.}\ \bibnamefont {Cheung}}, \bibinfo {author} {\bibfnamefont {C.}~\bibnamefont {O'Hara}}, \bibinfo {author} {\bibfnamefont {G.}~\bibnamefont {Moir}}, \bibinfo {author} {\bibfnamefont {M.}~\bibnamefont {Peardon}}, \bibinfo {author} {\bibfnamefont {S.~M.}\ \bibnamefont {Ryan}}, \bibinfo {author} {\bibfnamefont {C.~E.}\ \bibnamefont {Thomas}},\ and\ \bibinfo {author} {\bibfnamefont {D.}~\bibnamefont {Tims}} (\bibinfo {collaboration} {Hadron Spectrum}),\ }\bibfield  {title} {\bibinfo {title} {{Excited and exotic charmonium, $D_s$ and $D$ meson spectra for two light quark masses from lattice QCD}},\ }\href {https://doi.org/10.1007/JHEP12(2016)089} {\bibfield  {journal} {\bibinfo  {journal} {JHEP}\ }\textbf {\bibinfo {volume} {12}},\ \bibinfo {pages} {089}},\ \Eprint {https://arxiv.org/abs/1610.01073} {arXiv:1610.01073 [hep-lat]} \BibitemShut {NoStop}%
\bibitem [{\citenamefont {DeTar}\ \emph {et~al.}(2019)\citenamefont {DeTar}, \citenamefont {Kronfeld}, \citenamefont {Lee}, \citenamefont {Mohler},\ and\ \citenamefont {Simone}}]{DeTar:2018uko}%
  \BibitemOpen
  \bibfield  {author} {\bibinfo {author} {\bibfnamefont {C.}~\bibnamefont {DeTar}}, \bibinfo {author} {\bibfnamefont {A.~S.}\ \bibnamefont {Kronfeld}}, \bibinfo {author} {\bibfnamefont {S.-h.}\ \bibnamefont {Lee}}, \bibinfo {author} {\bibfnamefont {D.}~\bibnamefont {Mohler}},\ and\ \bibinfo {author} {\bibfnamefont {J.~N.}\ \bibnamefont {Simone}} (\bibinfo {collaboration} {Fermilab Lattice, MILC}),\ }\bibfield  {title} {\bibinfo {title} {{Splittings of low-lying charmonium masses at the physical point}},\ }\href {https://doi.org/10.1103/PhysRevD.99.034509} {\bibfield  {journal} {\bibinfo  {journal} {Phys.\ Rev.\ D}\ }\textbf {\bibinfo {volume} {99}},\ \bibinfo {pages} {034509} (\bibinfo {year} {2019})},\ \Eprint {https://arxiv.org/abs/1810.09983} {arXiv:1810.09983 [hep-lat]} \BibitemShut {NoStop}%
\bibitem [{\citenamefont {Hatton}\ \emph {et~al.}(2020)\citenamefont {Hatton}, \citenamefont {Davies}, \citenamefont {Galloway}, \citenamefont {Koponen}, \citenamefont {Lepage},\ and\ \citenamefont {Lytle}}]{Hatton}%
  \BibitemOpen
  \bibfield  {author} {\bibinfo {author} {\bibfnamefont {D.}~\bibnamefont {Hatton}}, \bibinfo {author} {\bibfnamefont {C.~T.~H.}\ \bibnamefont {Davies}}, \bibinfo {author} {\bibfnamefont {B.}~\bibnamefont {Galloway}}, \bibinfo {author} {\bibfnamefont {J.}~\bibnamefont {Koponen}}, \bibinfo {author} {\bibfnamefont {G.~P.}\ \bibnamefont {Lepage}},\ and\ \bibinfo {author} {\bibfnamefont {A.~T.}\ \bibnamefont {Lytle}} (\bibinfo {collaboration} {HPQCD Collaboration}),\ }\bibfield  {title} {\bibinfo {title} {Charmonium properties from lattice $\mathrm{QCD}+\text{QED}$: Hyperfine splitting, $j/\ensuremath{\psi}$ leptonic width, charm quark mass, and ${a}_{\ensuremath{\mu}}^{c}$},\ }\href {https://doi.org/10.1103/PhysRevD.102.054511} {\bibfield  {journal} {\bibinfo  {journal} {Phys. Rev. D}\ }\textbf {\bibinfo {volume} {102}},\ \bibinfo {pages} {054511} (\bibinfo {year} {2020})}\BibitemShut {NoStop}%
\bibitem [{\citenamefont {Close}(1979)}]{Close:1979bt}%
  \BibitemOpen
  \bibfield  {author} {\bibinfo {author} {\bibfnamefont {F.}~\bibnamefont {Close}},\ }\href@noop {} {\emph {\bibinfo {title} {{An Introduction to Quarks and Partons}}}}\ (\bibinfo {year} {1979})\BibitemShut {NoStop}%
\bibitem [{\citenamefont {Peardon}\ \emph {et~al.}(2009)\citenamefont {Peardon}, \citenamefont {Bulava}, \citenamefont {Foley}, \citenamefont {Morningstar}, \citenamefont {Dudek}, \citenamefont {Edwards}, \citenamefont {Joo}, \citenamefont {Lin}, \citenamefont {Richards},\ and\ \citenamefont {Juge}}]{Peardon:2009gh}%
  \BibitemOpen
  \bibfield  {author} {\bibinfo {author} {\bibfnamefont {M.}~\bibnamefont {Peardon}}, \bibinfo {author} {\bibfnamefont {J.}~\bibnamefont {Bulava}}, \bibinfo {author} {\bibfnamefont {J.}~\bibnamefont {Foley}}, \bibinfo {author} {\bibfnamefont {C.}~\bibnamefont {Morningstar}}, \bibinfo {author} {\bibfnamefont {J.}~\bibnamefont {Dudek}}, \bibinfo {author} {\bibfnamefont {R.~G.}\ \bibnamefont {Edwards}}, \bibinfo {author} {\bibfnamefont {B.}~\bibnamefont {Joo}}, \bibinfo {author} {\bibfnamefont {H.-W.}\ \bibnamefont {Lin}}, \bibinfo {author} {\bibfnamefont {D.~G.}\ \bibnamefont {Richards}},\ and\ \bibinfo {author} {\bibfnamefont {K.~J.}\ \bibnamefont {Juge}} (\bibinfo {collaboration} {Hadron Spectrum}),\ }\bibfield  {title} {\bibinfo {title} {{A Novel quark-field creation operator construction for hadronic physics in lattice QCD}},\ }\href {https://doi.org/10.1103/PhysRevD.80.054506} {\bibfield  {journal} {\bibinfo  {journal} {Phys.\ Rev.\ D}\ }\textbf {\bibinfo {volume} {80}},\ \bibinfo {pages} {054506}
  (\bibinfo {year} {2009})},\ \Eprint {https://arxiv.org/abs/0905.2160} {arXiv:0905.2160 [hep-lat]} \BibitemShut {NoStop}%
\bibitem [{\citenamefont {Knechtli}\ \emph {et~al.}(2022)\citenamefont {Knechtli}, \citenamefont {Korzec}, \citenamefont {Peardon},\ and\ \citenamefont {Urrea-Niño}}]{Knechtli2022}%
  \BibitemOpen
  \bibfield  {author} {\bibinfo {author} {\bibfnamefont {F.}~\bibnamefont {Knechtli}}, \bibinfo {author} {\bibfnamefont {T.}~\bibnamefont {Korzec}}, \bibinfo {author} {\bibfnamefont {M.}~\bibnamefont {Peardon}},\ and\ \bibinfo {author} {\bibfnamefont {J.~A.}\ \bibnamefont {Urrea-Niño}},\ }\bibfield  {title} {\bibinfo {title} {Optimizing creation operators for charmonium spectroscopy on the lattice},\ }\bibfield  {journal} {\bibinfo  {journal} {Physical Review D}\ }\textbf {\bibinfo {volume} {106}},\ \href {https://doi.org/10.1103/physrevd.106.034501} {10.1103/physrevd.106.034501} (\bibinfo {year} {2022})\BibitemShut {NoStop}%
\bibitem [{\citenamefont {Urrea-Ni\~no}\ \emph {et~al.}(2023)\citenamefont {Urrea-Ni\~no}, \citenamefont {Knechtli}, \citenamefont {Korzec},\ and\ \citenamefont {Peardon}}]{Urrea-Nino:2022gne}%
  \BibitemOpen
  \bibfield  {author} {\bibinfo {author} {\bibfnamefont {J.~A.}\ \bibnamefont {Urrea-Ni\~no}}, \bibinfo {author} {\bibfnamefont {F.}~\bibnamefont {Knechtli}}, \bibinfo {author} {\bibfnamefont {T.}~\bibnamefont {Korzec}},\ and\ \bibinfo {author} {\bibfnamefont {M.~J.}\ \bibnamefont {Peardon}},\ }\bibfield  {title} {\bibinfo {title} {{Optimized meson operators for charmonium spectroscopy and mixing with glueballs}},\ }\href {https://doi.org/10.22323/1.430.0087} {\bibfield  {journal} {\bibinfo  {journal} {PoS}\ }\textbf {\bibinfo {volume} {LATTICE2022}},\ \bibinfo {pages} {087} (\bibinfo {year} {2023})},\ \Eprint {https://arxiv.org/abs/2212.09404} {arXiv:2212.09404 [hep-lat]} \BibitemShut {NoStop}%
\bibitem [{\citenamefont {Ni\~no}\ \emph {et~al.}(2022)\citenamefont {Ni\~no}, \citenamefont {Knechtli}, \citenamefont {Korzec},\ and\ \citenamefont {Peardon}}]{Nino:2021klm}%
  \BibitemOpen
  \bibfield  {author} {\bibinfo {author} {\bibfnamefont {J.~A.~U.}\ \bibnamefont {Ni\~no}}, \bibinfo {author} {\bibfnamefont {F.}~\bibnamefont {Knechtli}}, \bibinfo {author} {\bibfnamefont {T.}~\bibnamefont {Korzec}},\ and\ \bibinfo {author} {\bibfnamefont {M.}~\bibnamefont {Peardon}},\ }\bibfield  {title} {\bibinfo {title} {{Optimizing distillation for charmonium and glueballs}},\ }\href {https://doi.org/10.22323/1.396.0314} {\bibfield  {journal} {\bibinfo  {journal} {PoS}\ }\textbf {\bibinfo {volume} {LATTICE2021}},\ \bibinfo {pages} {314} (\bibinfo {year} {2022})},\ \Eprint {https://arxiv.org/abs/2112.01964} {arXiv:2112.01964 [hep-lat]} \BibitemShut {NoStop}%
\bibitem [{\citenamefont {Dudek}\ \emph {et~al.}(2011)\citenamefont {Dudek}, \citenamefont {Edwards}, \citenamefont {Joo}, \citenamefont {Peardon}, \citenamefont {Richards},\ and\ \citenamefont {Thomas}}]{Dudek:2011tt}%
  \BibitemOpen
  \bibfield  {author} {\bibinfo {author} {\bibfnamefont {J.~J.}\ \bibnamefont {Dudek}}, \bibinfo {author} {\bibfnamefont {R.~G.}\ \bibnamefont {Edwards}}, \bibinfo {author} {\bibfnamefont {B.}~\bibnamefont {Joo}}, \bibinfo {author} {\bibfnamefont {M.~J.}\ \bibnamefont {Peardon}}, \bibinfo {author} {\bibfnamefont {D.~G.}\ \bibnamefont {Richards}},\ and\ \bibinfo {author} {\bibfnamefont {C.~E.}\ \bibnamefont {Thomas}},\ }\bibfield  {title} {\bibinfo {title} {{Isoscalar meson spectroscopy from lattice QCD}},\ }\href {https://doi.org/10.1103/PhysRevD.83.111502} {\bibfield  {journal} {\bibinfo  {journal} {Phys.\ Rev.\ D}\ }\textbf {\bibinfo {volume} {83}},\ \bibinfo {pages} {111502} (\bibinfo {year} {2011})},\ \Eprint {https://arxiv.org/abs/1102.4299} {arXiv:1102.4299 [hep-lat]} \BibitemShut {NoStop}%
\bibitem [{\citenamefont {Zhang}\ \emph {et~al.}(2022{\natexlab{a}})\citenamefont {Zhang}, \citenamefont {Sun}, \citenamefont {Chen}, \citenamefont {Gong}, \citenamefont {Gui},\ and\ \citenamefont {Liu}}]{Zhang2022}%
  \BibitemOpen
  \bibfield  {author} {\bibinfo {author} {\bibfnamefont {R.}~\bibnamefont {Zhang}}, \bibinfo {author} {\bibfnamefont {W.}~\bibnamefont {Sun}}, \bibinfo {author} {\bibfnamefont {Y.}~\bibnamefont {Chen}}, \bibinfo {author} {\bibfnamefont {M.}~\bibnamefont {Gong}}, \bibinfo {author} {\bibfnamefont {L.-C.}\ \bibnamefont {Gui}},\ and\ \bibinfo {author} {\bibfnamefont {Z.}~\bibnamefont {Liu}},\ }\bibfield  {title} {\bibinfo {title} {The glueball content of $\eta$},\ }\href {https://doi.org/10.1016/j.physletb.2022.136960} {\bibfield  {journal} {\bibinfo  {journal} {Physics Letters B}\ }\textbf {\bibinfo {volume} {827}},\ \bibinfo {pages} {136960} (\bibinfo {year} {2022}{\natexlab{a}})}\BibitemShut {NoStop}%
\bibitem [{\citenamefont {Jiang}\ \emph {et~al.}(2023)\citenamefont {Jiang}, \citenamefont {Sun}, \citenamefont {Chen}, \citenamefont {Chen}, \citenamefont {Gong}, \citenamefont {Liu},\ and\ \citenamefont {Zhang}}]{Jiang2023}%
  \BibitemOpen
  \bibfield  {author} {\bibinfo {author} {\bibfnamefont {X.}~\bibnamefont {Jiang}}, \bibinfo {author} {\bibfnamefont {W.}~\bibnamefont {Sun}}, \bibinfo {author} {\bibfnamefont {F.}~\bibnamefont {Chen}}, \bibinfo {author} {\bibfnamefont {Y.}~\bibnamefont {Chen}}, \bibinfo {author} {\bibfnamefont {M.}~\bibnamefont {Gong}}, \bibinfo {author} {\bibfnamefont {Z.}~\bibnamefont {Liu}},\ and\ \bibinfo {author} {\bibfnamefont {R.}~\bibnamefont {Zhang}},\ }\bibfield  {title} {\bibinfo {title} {$\ensuremath{\eta}$-glueball mixing from ${N}_{f}=2$ lattice qcd},\ }\href {https://doi.org/10.1103/PhysRevD.107.094510} {\bibfield  {journal} {\bibinfo  {journal} {Phys. Rev. D}\ }\textbf {\bibinfo {volume} {107}},\ \bibinfo {pages} {094510} (\bibinfo {year} {2023})}\BibitemShut {NoStop}%
\bibitem [{\citenamefont {Levkova}\ and\ \citenamefont {DeTar}(2011)}]{Levkova}%
  \BibitemOpen
  \bibfield  {author} {\bibinfo {author} {\bibfnamefont {L.}~\bibnamefont {Levkova}}\ and\ \bibinfo {author} {\bibfnamefont {C.}~\bibnamefont {DeTar}},\ }\bibfield  {title} {\bibinfo {title} {Charm annihilation effects on the hyperfine splitting in charmonium},\ }\href {https://doi.org/10.1103/PhysRevD.83.074504} {\bibfield  {journal} {\bibinfo  {journal} {Phys. Rev. D}\ }\textbf {\bibinfo {volume} {83}},\ \bibinfo {pages} {074504} (\bibinfo {year} {2011})}\BibitemShut {NoStop}%
\bibitem [{\citenamefont {McNeile}\ and\ \citenamefont {Michael}(2004)}]{McNeile}%
  \BibitemOpen
  \bibfield  {author} {\bibinfo {author} {\bibfnamefont {C.}~\bibnamefont {McNeile}}\ and\ \bibinfo {author} {\bibfnamefont {C.}~\bibnamefont {Michael}} (\bibinfo {collaboration} {UKQCD Collaboration}),\ }\bibfield  {title} {\bibinfo {title} {Estimate of the flavor singlet contributions to the hyperfine splitting in charmonium},\ }\href {https://doi.org/10.1103/PhysRevD.70.034506} {\bibfield  {journal} {\bibinfo  {journal} {Phys. Rev. D}\ }\textbf {\bibinfo {volume} {70}},\ \bibinfo {pages} {034506} (\bibinfo {year} {2004})}\BibitemShut {NoStop}%
\bibitem [{\citenamefont {de~Forcrand}\ \emph {et~al.}(2004)\citenamefont {de~Forcrand}, \citenamefont {Pérez}, \citenamefont {Matsufuru}, \citenamefont {Nakamura}, \citenamefont {Pushkina}, \citenamefont {Stamatescu}, \citenamefont {Takaishi},\ and\ \citenamefont {Umeda}}]{Forcrand}%
  \BibitemOpen
  \bibfield  {author} {\bibinfo {author} {\bibfnamefont {P.}~\bibnamefont {de~Forcrand}}, \bibinfo {author} {\bibfnamefont {M.~G.}\ \bibnamefont {Pérez}}, \bibinfo {author} {\bibfnamefont {H.}~\bibnamefont {Matsufuru}}, \bibinfo {author} {\bibfnamefont {A.}~\bibnamefont {Nakamura}}, \bibinfo {author} {\bibfnamefont {I.}~\bibnamefont {Pushkina}}, \bibinfo {author} {\bibfnamefont {I.-O.}\ \bibnamefont {Stamatescu}}, \bibinfo {author} {\bibfnamefont {T.}~\bibnamefont {Takaishi}},\ and\ \bibinfo {author} {\bibfnamefont {T.}~\bibnamefont {Umeda}},\ }\bibfield  {title} {\bibinfo {title} {Contribution of disconnected diagrams to the hyperfine splitting of charmonium},\ }\href {https://doi.org/10.1088/1126-6708/2004/08/004} {\bibfield  {journal} {\bibinfo  {journal} {Journal of High Energy Physics}\ }\textbf {\bibinfo {volume} {2004}},\ \bibinfo {pages} {004} (\bibinfo {year} {2004})}\BibitemShut {NoStop}%
\bibitem [{\citenamefont {Lüscher}\ and\ \citenamefont {Wolff}(1990)}]{Luscher:1990ck}%
  \BibitemOpen
  \bibfield  {author} {\bibinfo {author} {\bibfnamefont {M.}~\bibnamefont {Lüscher}}\ and\ \bibinfo {author} {\bibfnamefont {U.}~\bibnamefont {Wolff}},\ }\bibfield  {title} {\bibinfo {title} {{How to Calculate the Elastic Scattering Matrix in Two-dimensional Quantum Field Theories by Numerical Simulation}},\ }\href {https://doi.org/10.1016/0550-3213(90)90540-T} {\bibfield  {journal} {\bibinfo  {journal} {Nucl. Phys. B}\ }\textbf {\bibinfo {volume} {339}},\ \bibinfo {pages} {222} (\bibinfo {year} {1990})}\BibitemShut {NoStop}%
\bibitem [{\citenamefont {Blossier}\ \emph {et~al.}(2009)\citenamefont {Blossier}, \citenamefont {Della~Morte}, \citenamefont {von Hippel}, \citenamefont {Mendes},\ and\ \citenamefont {Sommer}}]{Blossier:2009kd}%
  \BibitemOpen
  \bibfield  {author} {\bibinfo {author} {\bibfnamefont {B.}~\bibnamefont {Blossier}}, \bibinfo {author} {\bibfnamefont {M.}~\bibnamefont {Della~Morte}}, \bibinfo {author} {\bibfnamefont {G.}~\bibnamefont {von Hippel}}, \bibinfo {author} {\bibfnamefont {T.}~\bibnamefont {Mendes}},\ and\ \bibinfo {author} {\bibfnamefont {R.}~\bibnamefont {Sommer}},\ }\bibfield  {title} {\bibinfo {title} {{On the generalized eigenvalue method for energies and matrix elements in lattice field theory}},\ }\href {https://doi.org/10.1088/1126-6708/2009/04/094} {\bibfield  {journal} {\bibinfo  {journal} {JHEP}\ }\textbf {\bibinfo {volume} {04}},\ \bibinfo {pages} {094}},\ \Eprint {https://arxiv.org/abs/0902.1265} {arXiv:0902.1265 [hep-lat]} \BibitemShut {NoStop}%
\bibitem [{\citenamefont {Balog}\ \emph {et~al.}(1999)\citenamefont {Balog}, \citenamefont {Niedermaier}, \citenamefont {Niedermayer}, \citenamefont {Patrascioiu}, \citenamefont {Seiler},\ and\ \citenamefont {Weisz}}]{Balog}%
  \BibitemOpen
  \bibfield  {author} {\bibinfo {author} {\bibfnamefont {J.}~\bibnamefont {Balog}}, \bibinfo {author} {\bibfnamefont {M.}~\bibnamefont {Niedermaier}}, \bibinfo {author} {\bibfnamefont {F.}~\bibnamefont {Niedermayer}}, \bibinfo {author} {\bibfnamefont {A.}~\bibnamefont {Patrascioiu}}, \bibinfo {author} {\bibfnamefont {E.}~\bibnamefont {Seiler}},\ and\ \bibinfo {author} {\bibfnamefont {P.}~\bibnamefont {Weisz}},\ }\bibfield  {title} {\bibinfo {title} {Comparison of the o(3) bootstrap \ensuremath{\sigma} model with lattice regularization at low energies},\ }\href {https://doi.org/10.1103/PhysRevD.60.094508} {\bibfield  {journal} {\bibinfo  {journal} {Phys. Rev. D}\ }\textbf {\bibinfo {volume} {60}},\ \bibinfo {pages} {094508} (\bibinfo {year} {1999})}\BibitemShut {NoStop}%
\bibitem [{\citenamefont {Niedermayer}\ \emph {et~al.}(2001)\citenamefont {Niedermayer}, \citenamefont {Rüfenacht},\ and\ \citenamefont {Wenger}}]{Niedermayer}%
  \BibitemOpen
  \bibfield  {author} {\bibinfo {author} {\bibfnamefont {F.}~\bibnamefont {Niedermayer}}, \bibinfo {author} {\bibfnamefont {P.}~\bibnamefont {Rüfenacht}},\ and\ \bibinfo {author} {\bibfnamefont {U.}~\bibnamefont {Wenger}},\ }\bibfield  {title} {\bibinfo {title} {Fixed point gauge actions with fat links: scaling and glueballs},\ }\href {https://doi.org/https://doi.org/10.1016/S0550-3213(00)00731-8} {\bibfield  {journal} {\bibinfo  {journal} {Nuclear Physics B}\ }\textbf {\bibinfo {volume} {597}},\ \bibinfo {pages} {413} (\bibinfo {year} {2001})}\BibitemShut {NoStop}%
\bibitem [{\citenamefont {Bali}\ \emph {et~al.}(2005)\citenamefont {Bali}, \citenamefont {Neff}, \citenamefont {Duessel}, \citenamefont {Lippert},\ and\ \citenamefont {Schilling}}]{Bali:2005fu}%
  \BibitemOpen
  \bibfield  {author} {\bibinfo {author} {\bibfnamefont {G.~S.}\ \bibnamefont {Bali}}, \bibinfo {author} {\bibfnamefont {H.}~\bibnamefont {Neff}}, \bibinfo {author} {\bibfnamefont {T.}~\bibnamefont {Duessel}}, \bibinfo {author} {\bibfnamefont {T.}~\bibnamefont {Lippert}},\ and\ \bibinfo {author} {\bibfnamefont {K.}~\bibnamefont {Schilling}} (\bibinfo {collaboration} {SESAM}),\ }\bibfield  {title} {\bibinfo {title} {{Observation of string breaking in QCD}},\ }\href {https://doi.org/10.1103/PhysRevD.71.114513} {\bibfield  {journal} {\bibinfo  {journal} {Phys. Rev. D}\ }\textbf {\bibinfo {volume} {71}},\ \bibinfo {pages} {114513} (\bibinfo {year} {2005})},\ \Eprint {https://arxiv.org/abs/hep-lat/0505012} {arXiv:hep-lat/0505012} \BibitemShut {NoStop}%
\bibitem [{\citenamefont {Bali}\ \emph {et~al.}(2015)\citenamefont {Bali}, \citenamefont {Collins}, \citenamefont {D\"urr},\ and\ \citenamefont {Kanamori}}]{BaliArtifacts}%
  \BibitemOpen
  \bibfield  {author} {\bibinfo {author} {\bibfnamefont {G.~S.}\ \bibnamefont {Bali}}, \bibinfo {author} {\bibfnamefont {S.}~\bibnamefont {Collins}}, \bibinfo {author} {\bibfnamefont {S.}~\bibnamefont {D\"urr}},\ and\ \bibinfo {author} {\bibfnamefont {I.}~\bibnamefont {Kanamori}},\ }\bibfield  {title} {\bibinfo {title} {${D}_{s}\ensuremath{\rightarrow}\ensuremath{\eta},\ensuremath{\eta}\ensuremath{'}$ semileptonic decay form factors with disconnected quark loop contributions},\ }\href {https://doi.org/10.1103/PhysRevD.91.014503} {\bibfield  {journal} {\bibinfo  {journal} {Phys. Rev. D}\ }\textbf {\bibinfo {volume} {91}},\ \bibinfo {pages} {014503} (\bibinfo {year} {2015})}\BibitemShut {NoStop}%
\bibitem [{\citenamefont {Dimopoulos}\ \emph {et~al.}(2019)\citenamefont {Dimopoulos}, \citenamefont {Helmes}, \citenamefont {Jost}, \citenamefont {Knippschild}, \citenamefont {Kostrzewa}, \citenamefont {Liu}, \citenamefont {Ottnad}, \citenamefont {Petschlies}, \citenamefont {Urbach}, \citenamefont {Wenger},\ and\ \citenamefont {Werner}}]{UrbachArtifacts}%
  \BibitemOpen
  \bibfield  {author} {\bibinfo {author} {\bibfnamefont {P.}~\bibnamefont {Dimopoulos}}, \bibinfo {author} {\bibfnamefont {C.}~\bibnamefont {Helmes}}, \bibinfo {author} {\bibfnamefont {C.}~\bibnamefont {Jost}}, \bibinfo {author} {\bibfnamefont {B.}~\bibnamefont {Knippschild}}, \bibinfo {author} {\bibfnamefont {B.}~\bibnamefont {Kostrzewa}}, \bibinfo {author} {\bibfnamefont {L.}~\bibnamefont {Liu}}, \bibinfo {author} {\bibfnamefont {K.}~\bibnamefont {Ottnad}}, \bibinfo {author} {\bibfnamefont {M.}~\bibnamefont {Petschlies}}, \bibinfo {author} {\bibfnamefont {C.}~\bibnamefont {Urbach}}, \bibinfo {author} {\bibfnamefont {U.}~\bibnamefont {Wenger}},\ and\ \bibinfo {author} {\bibfnamefont {M.}~\bibnamefont {Werner}} (\bibinfo {collaboration} {ETM Collaboration}),\ }\bibfield  {title} {\bibinfo {title} {Topological susceptibility and ${\ensuremath{\eta}}^{\ensuremath{'}}$ meson mass from ${N}_{f}=2$ lattice qcd at the physical point},\ }\href {https://doi.org/10.1103/PhysRevD.99.034511} {\bibfield  {journal}
  {\bibinfo  {journal} {Phys. Rev. D}\ }\textbf {\bibinfo {volume} {99}},\ \bibinfo {pages} {034511} (\bibinfo {year} {2019})}\BibitemShut {NoStop}%
\bibitem [{\citenamefont {Aoki}\ \emph {et~al.}(2007)\citenamefont {Aoki}, \citenamefont {Fukaya}, \citenamefont {Hashimoto},\ and\ \citenamefont {Onogi}}]{AokiArtifacts}%
  \BibitemOpen
  \bibfield  {author} {\bibinfo {author} {\bibfnamefont {S.}~\bibnamefont {Aoki}}, \bibinfo {author} {\bibfnamefont {H.}~\bibnamefont {Fukaya}}, \bibinfo {author} {\bibfnamefont {S.}~\bibnamefont {Hashimoto}},\ and\ \bibinfo {author} {\bibfnamefont {T.}~\bibnamefont {Onogi}},\ }\bibfield  {title} {\bibinfo {title} {Finite volume qcd at fixed topological charge},\ }\href {https://doi.org/10.1103/PhysRevD.76.054508} {\bibfield  {journal} {\bibinfo  {journal} {Phys. Rev. D}\ }\textbf {\bibinfo {volume} {76}},\ \bibinfo {pages} {054508} (\bibinfo {year} {2007})}\BibitemShut {NoStop}%
\bibitem [{\citenamefont {Ottnad}\ and\ \citenamefont {Urbach}(2018)}]{OttnadArtifacts}%
  \BibitemOpen
  \bibfield  {author} {\bibinfo {author} {\bibfnamefont {K.}~\bibnamefont {Ottnad}}\ and\ \bibinfo {author} {\bibfnamefont {C.}~\bibnamefont {Urbach}} (\bibinfo {collaboration} {ETM Collaboration}),\ }\bibfield  {title} {\bibinfo {title} {Flavor-singlet meson decay constants from ${N}_{f}=2+1+1$ twisted mass lattice qcd},\ }\href {https://doi.org/10.1103/PhysRevD.97.054508} {\bibfield  {journal} {\bibinfo  {journal} {Phys. Rev. D}\ }\textbf {\bibinfo {volume} {97}},\ \bibinfo {pages} {054508} (\bibinfo {year} {2018})}\BibitemShut {NoStop}%
\bibitem [{\citenamefont {Joswig}\ \emph {et~al.}(2023)\citenamefont {Joswig}, \citenamefont {Kuberski}, \citenamefont {Kuhlmann},\ and\ \citenamefont {Neuendorf}}]{Joswig2023}%
  \BibitemOpen
  \bibfield  {author} {\bibinfo {author} {\bibfnamefont {F.}~\bibnamefont {Joswig}}, \bibinfo {author} {\bibfnamefont {S.}~\bibnamefont {Kuberski}}, \bibinfo {author} {\bibfnamefont {J.~T.}\ \bibnamefont {Kuhlmann}},\ and\ \bibinfo {author} {\bibfnamefont {J.}~\bibnamefont {Neuendorf}},\ }\bibfield  {title} {\bibinfo {title} {pyerrors: A python framework for error analysis of monte carlo data},\ }\href {https://doi.org/10.1016/j.cpc.2023.108750} {\bibfield  {journal} {\bibinfo  {journal} {Computer Physics Communications}\ }\textbf {\bibinfo {volume} {288}},\ \bibinfo {pages} {108750} (\bibinfo {year} {2023})}\BibitemShut {NoStop}%
\bibitem [{\citenamefont {Wolff}(2004)}]{Wolff2004}%
  \BibitemOpen
  \bibfield  {author} {\bibinfo {author} {\bibfnamefont {U.}~\bibnamefont {Wolff}},\ }\bibfield  {title} {\bibinfo {title} {Monte carlo errors with less errors},\ }\href {https://doi.org/10.1016/s0010-4655(03)00467-3} {\bibfield  {journal} {\bibinfo  {journal} {Computer Physics Communications}\ }\textbf {\bibinfo {volume} {156}},\ \bibinfo {pages} {143–153} (\bibinfo {year} {2004})}\BibitemShut {NoStop}%
\bibitem [{\citenamefont {Wolff}(2007)}]{Wolff2007}%
  \BibitemOpen
  \bibfield  {author} {\bibinfo {author} {\bibfnamefont {U.}~\bibnamefont {Wolff}},\ }\bibfield  {title} {\bibinfo {title} {Erratum to “monte carlo errors with less errors” [comput. phys. comm. 156 (2004) 143–153]},\ }\href {https://doi.org/10.1016/j.cpc.2006.12.001} {\bibfield  {journal} {\bibinfo  {journal} {Computer Physics Communications}\ }\textbf {\bibinfo {volume} {176}},\ \bibinfo {pages} {383} (\bibinfo {year} {2007})}\BibitemShut {NoStop}%
\bibitem [{\citenamefont {Schaefer}\ \emph {et~al.}(2011)\citenamefont {Schaefer}, \citenamefont {Sommer},\ and\ \citenamefont {Virotta}}]{Schaefer2011}%
  \BibitemOpen
  \bibfield  {author} {\bibinfo {author} {\bibfnamefont {S.}~\bibnamefont {Schaefer}}, \bibinfo {author} {\bibfnamefont {R.}~\bibnamefont {Sommer}},\ and\ \bibinfo {author} {\bibfnamefont {F.}~\bibnamefont {Virotta}},\ }\bibfield  {title} {\bibinfo {title} {Critical slowing down and error analysis in lattice qcd simulations},\ }\href {https://doi.org/10.1016/j.nuclphysb.2010.11.020} {\bibfield  {journal} {\bibinfo  {journal} {Nuclear Physics B}\ }\textbf {\bibinfo {volume} {845}},\ \bibinfo {pages} {93–119} (\bibinfo {year} {2011})}\BibitemShut {NoStop}%
\bibitem [{\citenamefont {Ramos}(2019)}]{Ramos2019}%
  \BibitemOpen
  \bibfield  {author} {\bibinfo {author} {\bibfnamefont {A.}~\bibnamefont {Ramos}},\ }\bibfield  {title} {\bibinfo {title} {Automatic differentiation for error analysis of monte carlo data},\ }\href {https://doi.org/10.1016/j.cpc.2018.12.020} {\bibfield  {journal} {\bibinfo  {journal} {Computer Physics Communications}\ }\textbf {\bibinfo {volume} {238}},\ \bibinfo {pages} {19–35} (\bibinfo {year} {2019})}\BibitemShut {NoStop}%
\bibitem [{\citenamefont {Dudek}\ \emph {et~al.}(2008)\citenamefont {Dudek}, \citenamefont {Edwards}, \citenamefont {Mathur},\ and\ \citenamefont {Richards}}]{Dudek2008}%
  \BibitemOpen
  \bibfield  {author} {\bibinfo {author} {\bibfnamefont {J.~J.}\ \bibnamefont {Dudek}}, \bibinfo {author} {\bibfnamefont {R.~G.}\ \bibnamefont {Edwards}}, \bibinfo {author} {\bibfnamefont {N.}~\bibnamefont {Mathur}},\ and\ \bibinfo {author} {\bibfnamefont {D.~G.}\ \bibnamefont {Richards}},\ }\bibfield  {title} {\bibinfo {title} {Charmonium excited state spectrum in lattice qcd},\ }\bibfield  {journal} {\bibinfo  {journal} {Physical Review D}\ }\textbf {\bibinfo {volume} {77}},\ \href {https://doi.org/10.1103/physrevd.77.034501} {10.1103/physrevd.77.034501} (\bibinfo {year} {2008})\BibitemShut {NoStop}%
\bibitem [{\citenamefont {Morningstar}\ and\ \citenamefont {Peardon}(1999)}]{Morningstar:1999rf}%
  \BibitemOpen
  \bibfield  {author} {\bibinfo {author} {\bibfnamefont {C.~J.}\ \bibnamefont {Morningstar}}\ and\ \bibinfo {author} {\bibfnamefont {M.~J.}\ \bibnamefont {Peardon}},\ }\bibfield  {title} {\bibinfo {title} {{The Glueball spectrum from an anisotropic lattice study}},\ }\href {https://doi.org/10.1103/PhysRevD.60.034509} {\bibfield  {journal} {\bibinfo  {journal} {Phys.\ Rev.\ D}\ }\textbf {\bibinfo {volume} {60}},\ \bibinfo {pages} {034509} (\bibinfo {year} {1999})},\ \Eprint {https://arxiv.org/abs/hep-lat/9901004} {arXiv:hep-lat/9901004} \BibitemShut {NoStop}%
\bibitem [{\citenamefont {{R. H\"ollwieser, F. Knechtli and T. Korzec}}(2020)}]{Hollwieser:2020qri}%
  \BibitemOpen
  \bibfield  {author} {\bibinfo {author} {\bibnamefont {{R. H\"ollwieser, F. Knechtli and T. Korzec}}} (\bibinfo {collaboration} {ALPHA}),\ }\bibfield  {title} {\bibinfo {title} {{Scale setting for $N_f=3+1$ QCD}},\ }\href {https://doi.org/10.1140/epjc/s10052-020-7889-7} {\bibfield  {journal} {\bibinfo  {journal} {Eur. Phys. J. C}\ }\textbf {\bibinfo {volume} {80}},\ \bibinfo {pages} {349} (\bibinfo {year} {2020})},\ \Eprint {https://arxiv.org/abs/2002.02866} {arXiv:2002.02866 [hep-lat]} \BibitemShut {NoStop}%
\bibitem [{\citenamefont {H\"ollwieser}\ \emph {et~al.}(2019)\citenamefont {H\"ollwieser}, \citenamefont {Knechtli},\ and\ \citenamefont {Korzec}}]{Hollwieser:2019kuc}%
  \BibitemOpen
  \bibfield  {author} {\bibinfo {author} {\bibfnamefont {R.}~\bibnamefont {H\"ollwieser}}, \bibinfo {author} {\bibfnamefont {F.}~\bibnamefont {Knechtli}},\ and\ \bibinfo {author} {\bibfnamefont {T.}~\bibnamefont {Korzec}},\ }\bibfield  {title} {\bibinfo {title} {{Scale setting for QCD with $N_f=3+1$ dynamical quarks}},\ }\href {https://doi.org/10.22323/1.363.0025} {\bibfield  {journal} {\bibinfo  {journal} {PoS}\ }\textbf {\bibinfo {volume} {LATTICE2019}},\ \bibinfo {pages} {025} (\bibinfo {year} {2019})},\ \Eprint {https://arxiv.org/abs/1907.04309} {arXiv:1907.04309 [hep-lat]} \BibitemShut {NoStop}%
\bibitem [{\citenamefont {Fritzsch}\ \emph {et~al.}(2018)\citenamefont {Fritzsch}, \citenamefont {Sommer}, \citenamefont {Stollenwerk},\ and\ \citenamefont {Wolff}}]{Fritzsch:2018kjg}%
  \BibitemOpen
  \bibfield  {author} {\bibinfo {author} {\bibfnamefont {P.}~\bibnamefont {Fritzsch}}, \bibinfo {author} {\bibfnamefont {R.}~\bibnamefont {Sommer}}, \bibinfo {author} {\bibfnamefont {F.}~\bibnamefont {Stollenwerk}},\ and\ \bibinfo {author} {\bibfnamefont {U.}~\bibnamefont {Wolff}} (\bibinfo {collaboration} {ALPHA}),\ }\bibfield  {title} {\bibinfo {title} {{Symanzik Improvement with Dynamical Charm: A 3+1 Scheme for Wilson Quarks}},\ }\href {https://doi.org/10.1007/JHEP06(2018)025} {\bibfield  {journal} {\bibinfo  {journal} {JHEP}\ }\textbf {\bibinfo {volume} {06}},\ \bibinfo {pages} {025}},\ \Eprint {https://arxiv.org/abs/1805.01661} {arXiv:1805.01661 [hep-lat]} \BibitemShut {NoStop}%
\bibitem [{\citenamefont {Sch\"afers}\ \emph {et~al.}(2025{\natexlab{a}})\citenamefont {Sch\"afers}, \citenamefont {Finkenrath}, \citenamefont {G\"unther},\ and\ \citenamefont {Knechtli}}]{Schafers:2024teg}%
  \BibitemOpen
  \bibfield  {author} {\bibinfo {author} {\bibfnamefont {K.}~\bibnamefont {Sch\"afers}}, \bibinfo {author} {\bibfnamefont {J.}~\bibnamefont {Finkenrath}}, \bibinfo {author} {\bibfnamefont {M.}~\bibnamefont {G\"unther}},\ and\ \bibinfo {author} {\bibfnamefont {F.}~\bibnamefont {Knechtli}},\ }\bibfield  {title} {\bibinfo {title} {{Hessian-free force-gradient integrators}},\ }\href {https://doi.org/10.1016/j.cpc.2024.109478} {\bibfield  {journal} {\bibinfo  {journal} {Comput. Phys. Commun.}\ }\textbf {\bibinfo {volume} {309}},\ \bibinfo {pages} {109478} (\bibinfo {year} {2025}{\natexlab{a}})},\ \Eprint {https://arxiv.org/abs/2403.10370} {arXiv:2403.10370 [math.NA]} \BibitemShut {NoStop}%
\bibitem [{\citenamefont {Sch\"afers}\ \emph {et~al.}(2025{\natexlab{b}})\citenamefont {Sch\"afers}, \citenamefont {Finkenrath}, \citenamefont {G\"unther},\ and\ \citenamefont {Knechtli}}]{Schafers:2025mgt}%
  \BibitemOpen
  \bibfield  {author} {\bibinfo {author} {\bibfnamefont {K.}~\bibnamefont {Sch\"afers}}, \bibinfo {author} {\bibfnamefont {J.}~\bibnamefont {Finkenrath}}, \bibinfo {author} {\bibfnamefont {M.}~\bibnamefont {G\"unther}},\ and\ \bibinfo {author} {\bibfnamefont {F.}~\bibnamefont {Knechtli}},\ }\bibfield  {title} {\bibinfo {title} {{Numerical stability of force-gradient integrators and their Hessian-free variants in lattice QCD simulations}},\ }\href@noop {} {\  (\bibinfo {year} {2025}{\natexlab{b}})},\ \Eprint {https://arxiv.org/abs/2506.08813} {arXiv:2506.08813 [hep-lat]} \BibitemShut {NoStop}%
\bibitem [{\citenamefont {L{\"u}scher}\ and\ \citenamefont {Schaefer}(2013)}]{Luscher:2012av}%
  \BibitemOpen
  \bibfield  {author} {\bibinfo {author} {\bibfnamefont {M.}~\bibnamefont {L{\"u}scher}}\ and\ \bibinfo {author} {\bibfnamefont {S.}~\bibnamefont {Schaefer}},\ }\bibfield  {title} {\bibinfo {title} {{Lattice QCD with open boundary conditions and twisted-mass reweighting}},\ }\href {https://doi.org/10.1016/j.cpc.2012.10.003} {\bibfield  {journal} {\bibinfo  {journal} {Comput.\ Phys.\ Commun.}\ }\textbf {\bibinfo {volume} {184}},\ \bibinfo {pages} {519} (\bibinfo {year} {2013})},\ \Eprint {https://arxiv.org/abs/1206.2809} {arXiv:1206.2809 [hep-lat]} \BibitemShut {NoStop}%
\bibitem [{\citenamefont {Navas}\ \emph {et~al.}(2024)\citenamefont {Navas} \emph {et~al.}}]{ParticleDataGroup:2024cfk}%
  \BibitemOpen
  \bibfield  {author} {\bibinfo {author} {\bibfnamefont {S.}~\bibnamefont {Navas}} \emph {et~al.} (\bibinfo {collaboration} {Particle Data Group}),\ }\bibfield  {title} {\bibinfo {title} {{Review of particle physics}},\ }\href {https://doi.org/10.1103/PhysRevD.110.030001} {\bibfield  {journal} {\bibinfo  {journal} {Phys. Rev. D}\ }\textbf {\bibinfo {volume} {110}},\ \bibinfo {pages} {030001} (\bibinfo {year} {2024})}\BibitemShut {NoStop}%
\bibitem [{\citenamefont {no}\ \emph {et~al.}(2025)\citenamefont {no}, \citenamefont {Finkenrath}, \citenamefont {Höllwieser}, \citenamefont {Knechtli}, \citenamefont {Korzec},\ and\ \citenamefont {Peardon}}]{urreanino2025flavormixingcharmoniumlight}%
  \BibitemOpen
  \bibfield  {author} {\bibinfo {author} {\bibfnamefont {J.~A. U.-N.}\ \bibnamefont {no}}, \bibinfo {author} {\bibfnamefont {J.}~\bibnamefont {Finkenrath}}, \bibinfo {author} {\bibfnamefont {R.}~\bibnamefont {Höllwieser}}, \bibinfo {author} {\bibfnamefont {F.}~\bibnamefont {Knechtli}}, \bibinfo {author} {\bibfnamefont {T.}~\bibnamefont {Korzec}},\ and\ \bibinfo {author} {\bibfnamefont {M.}~\bibnamefont {Peardon}},\ }\href {https://arxiv.org/abs/2502.04977} {\bibinfo {title} {Flavor mixing in charmonium and light mesons with optimal distillation profiles}} (\bibinfo {year} {2025}),\ \Eprint {https://arxiv.org/abs/2502.04977} {arXiv:2502.04977 [hep-lat]} \BibitemShut {NoStop}%
\bibitem [{\citenamefont {Berg}\ and\ \citenamefont {Billoire}(1983)}]{Berg}%
  \BibitemOpen
  \bibfield  {author} {\bibinfo {author} {\bibfnamefont {B.}~\bibnamefont {Berg}}\ and\ \bibinfo {author} {\bibfnamefont {A.}~\bibnamefont {Billoire}},\ }\bibfield  {title} {\bibinfo {title} {Glueball spectroscopy in 4d {SU(3)} lattice gauge theory (i)},\ }\href {https://doi.org/https://doi.org/10.1016/0550-3213(83)90620-X} {\bibfield  {journal} {\bibinfo  {journal} {Nuclear Physics B}\ }\textbf {\bibinfo {volume} {221}},\ \bibinfo {pages} {109} (\bibinfo {year} {1983})}\BibitemShut {NoStop}%
\bibitem [{\citenamefont {Barca}\ \emph {et~al.}(2024)\citenamefont {Barca}, \citenamefont {Schaefer}, \citenamefont {Knechtli}, \citenamefont {Urrea-Ni\~no}, \citenamefont {Martins},\ and\ \citenamefont {Peardon}}]{Barca2024}%
  \BibitemOpen
  \bibfield  {author} {\bibinfo {author} {\bibfnamefont {L.}~\bibnamefont {Barca}}, \bibinfo {author} {\bibfnamefont {S.}~\bibnamefont {Schaefer}}, \bibinfo {author} {\bibfnamefont {F.}~\bibnamefont {Knechtli}}, \bibinfo {author} {\bibfnamefont {J.~A.}\ \bibnamefont {Urrea-Ni\~no}}, \bibinfo {author} {\bibfnamefont {S.}~\bibnamefont {Martins}},\ and\ \bibinfo {author} {\bibfnamefont {M.}~\bibnamefont {Peardon}},\ }\bibfield  {title} {\bibinfo {title} {Exponential error reduction for glueball calculations using a two-level algorithm in pure gauge theory},\ }\href {https://doi.org/10.1103/PhysRevD.110.054515} {\bibfield  {journal} {\bibinfo  {journal} {Phys. Rev. D}\ }\textbf {\bibinfo {volume} {110}},\ \bibinfo {pages} {054515} (\bibinfo {year} {2024})}\BibitemShut {NoStop}%
\bibitem [{\citenamefont {Athenodorou}\ \emph {et~al.}(2023)\citenamefont {Athenodorou}, \citenamefont {Finkenrath}, \citenamefont {Lantos},\ and\ \citenamefont {Teper}}]{Athenodorou2023}%
  \BibitemOpen
  \bibfield  {author} {\bibinfo {author} {\bibfnamefont {A.}~\bibnamefont {Athenodorou}}, \bibinfo {author} {\bibfnamefont {J.}~\bibnamefont {Finkenrath}}, \bibinfo {author} {\bibfnamefont {A.}~\bibnamefont {Lantos}},\ and\ \bibinfo {author} {\bibfnamefont {M.}~\bibnamefont {Teper}},\ }\href {https://arxiv.org/abs/2308.10054} {\bibinfo {title} {Glueball spectrum with four light dynamical fermions}} (\bibinfo {year} {2023}),\ \Eprint {https://arxiv.org/abs/2308.10054} {arXiv:2308.10054 [hep-lat]} \BibitemShut {NoStop}%
\bibitem [{\citenamefont {Albanese}\ \emph {et~al.}(1987)\citenamefont {Albanese} \emph {et~al.}}]{Albanese:1987ds}%
  \BibitemOpen
  \bibfield  {author} {\bibinfo {author} {\bibfnamefont {M.}~\bibnamefont {Albanese}} \emph {et~al.} (\bibinfo {collaboration} {APE}),\ }\bibfield  {title} {\bibinfo {title} {{Glueball Masses and String Tension in Lattice QCD}},\ }\href {https://doi.org/10.1016/0370-2693(87)91160-9} {\bibfield  {journal} {\bibinfo  {journal} {Phys. Lett. B}\ }\textbf {\bibinfo {volume} {192}},\ \bibinfo {pages} {163} (\bibinfo {year} {1987})}\BibitemShut {NoStop}%
\bibitem [{\citenamefont {McNeile}\ \emph {et~al.}(2001)\citenamefont {McNeile}, \citenamefont {Michael},\ and\ \citenamefont {Sharkey}}]{McNeil2}%
  \BibitemOpen
  \bibfield  {author} {\bibinfo {author} {\bibfnamefont {C.}~\bibnamefont {McNeile}}, \bibinfo {author} {\bibfnamefont {C.}~\bibnamefont {Michael}},\ and\ \bibinfo {author} {\bibfnamefont {K.~J.}\ \bibnamefont {Sharkey}} (\bibinfo {collaboration} {UKQCD Collaboration}),\ }\bibfield  {title} {\bibinfo {title} {Flavor singlet mesons in qcd},\ }\href {https://doi.org/10.1103/PhysRevD.65.014508} {\bibfield  {journal} {\bibinfo  {journal} {Phys. Rev. D}\ }\textbf {\bibinfo {volume} {65}},\ \bibinfo {pages} {014508} (\bibinfo {year} {2001})}\BibitemShut {NoStop}%
\bibitem [{\citenamefont {Follana}\ \emph {et~al.}(2007)\citenamefont {Follana}, \citenamefont {Mason}, \citenamefont {Davies}, \citenamefont {Hornbostel}, \citenamefont {Lepage}, \citenamefont {Shigemitsu}, \citenamefont {Trottier},\ and\ \citenamefont {Wong}}]{Follana}%
  \BibitemOpen
  \bibfield  {author} {\bibinfo {author} {\bibfnamefont {E.}~\bibnamefont {Follana}}, \bibinfo {author} {\bibfnamefont {Q.}~\bibnamefont {Mason}}, \bibinfo {author} {\bibfnamefont {C.}~\bibnamefont {Davies}}, \bibinfo {author} {\bibfnamefont {K.}~\bibnamefont {Hornbostel}}, \bibinfo {author} {\bibfnamefont {G.~P.}\ \bibnamefont {Lepage}}, \bibinfo {author} {\bibfnamefont {J.}~\bibnamefont {Shigemitsu}}, \bibinfo {author} {\bibfnamefont {H.}~\bibnamefont {Trottier}},\ and\ \bibinfo {author} {\bibfnamefont {K.}~\bibnamefont {Wong}},\ }\bibfield  {title} {\bibinfo {title} {Highly improved staggered quarks on the lattice with applications to charm physics},\ }\href {https://doi.org/10.1103/PhysRevD.75.054502} {\bibfield  {journal} {\bibinfo  {journal} {Phys. Rev. D}\ }\textbf {\bibinfo {volume} {75}},\ \bibinfo {pages} {054502} (\bibinfo {year} {2007})}\BibitemShut {NoStop}%
\bibitem [{\citenamefont {Bodwin}\ \emph {et~al.}(1995)\citenamefont {Bodwin}, \citenamefont {Braaten},\ and\ \citenamefont {Lepage}}]{Braaten}%
  \BibitemOpen
  \bibfield  {author} {\bibinfo {author} {\bibfnamefont {G.~T.}\ \bibnamefont {Bodwin}}, \bibinfo {author} {\bibfnamefont {E.}~\bibnamefont {Braaten}},\ and\ \bibinfo {author} {\bibfnamefont {G.~P.}\ \bibnamefont {Lepage}},\ }\bibfield  {title} {\bibinfo {title} {Rigorous qcd analysis of inclusive annihilation and production of heavy quarkonium},\ }\href {https://doi.org/10.1103/PhysRevD.51.1125} {\bibfield  {journal} {\bibinfo  {journal} {Phys. Rev. D}\ }\textbf {\bibinfo {volume} {51}},\ \bibinfo {pages} {1125} (\bibinfo {year} {1995})}\BibitemShut {NoStop}%
\bibitem [{\citenamefont {Zhang}\ \emph {et~al.}(2022{\natexlab{b}})\citenamefont {Zhang}, \citenamefont {Sun}, \citenamefont {Chen}, \citenamefont {Chen}, \citenamefont {Gong}, \citenamefont {Jiang},\ and\ \citenamefont {Liu}}]{Zhang_2022}%
  \BibitemOpen
  \bibfield  {author} {\bibinfo {author} {\bibfnamefont {R.}~\bibnamefont {Zhang}}, \bibinfo {author} {\bibfnamefont {W.}~\bibnamefont {Sun}}, \bibinfo {author} {\bibfnamefont {F.}~\bibnamefont {Chen}}, \bibinfo {author} {\bibfnamefont {Y.}~\bibnamefont {Chen}}, \bibinfo {author} {\bibfnamefont {M.}~\bibnamefont {Gong}}, \bibinfo {author} {\bibfnamefont {X.}~\bibnamefont {Jiang}},\ and\ \bibinfo {author} {\bibfnamefont {Z.}~\bibnamefont {Liu}},\ }\bibfield  {title} {\bibinfo {title} {Annihilation diagram contribution to charmonium masses *},\ }\href {https://doi.org/10.1088/1674-1137/ac3d8c} {\bibfield  {journal} {\bibinfo  {journal} {Chinese Physics C}\ }\textbf {\bibinfo {volume} {46}},\ \bibinfo {pages} {043102} (\bibinfo {year} {2022}{\natexlab{b}})}\BibitemShut {NoStop}%
\end{thebibliography}%

\end{document}